\documentclass[11pt]{article}

\usepackage[utf8]{inputenc}
\usepackage[T1]{fontenc}
\usepackage{amsmath, amssymb, amsthm}
\usepackage{mathtools}
\usepackage{graphicx}
\usepackage{booktabs}
\usepackage{natbib}
\usepackage{geometry}
\usepackage{setspace}
\usepackage{enumitem}
\usepackage[colorlinks=false,pdfborder={0 0 0}]{hyperref}
\usepackage{xcolor}
\usepackage{bm}             
\usepackage{authblk}        

\usepackage{etoolbox}

\theoremstyle{definition}

\theoremstyle{remark}

\title{\textbf{The Bayesian Way: \\ Uncertainty, Learning, and Statistical Reasoning}}
\author[1]{Juan Sosa\thanks{Email: \texttt{jcsosam@unal.edu.co}}}
\affil[1]{Departamento de Estadística, Universidad Nacional de Colombia}
\author[2]{Carlos A. Martínez\thanks{Email: \texttt{camartinezn@unal.edu.co}}}
\affil[2]{Departamento de Producción Animal, Universidad Nacional de Colombia}
\author[1]{Danna Cruz-Reyes\thanks{Email: \texttt{dlcruzr@unal.edu.co}}}
\date{}

\begin{document}

\maketitle

\begin{abstract}
This paper offers a comprehensive introduction to Bayesian inference, combining historical context, theoretical foundations, and core analytical examples. Beginning with Bayes’ theorem and the philosophical distinctions between Bayesian and frequentist approaches, we develop the inferential framework for estimation, interval construction, hypothesis testing, and prediction. Through canonical models, we illustrate how prior information and observed data are formally integrated to yield posterior distributions. We also explore key concepts including loss functions, credible intervals, Bayes factors, identifiability, and asymptotic behavior. While emphasizing analytical tractability in classical settings, we outline modern extensions that rely on simulation-based methods and discuss challenges related to prior specification and model evaluation. Though focused on foundational ideas, this paper sets the stage for applying Bayesian methods in contemporary domains such as hierarchical modeling, nonparametrics, and structured applications in time series, spatial data, networks, and political science. The goal is to provide a rigorous yet accessible entry point for students and researchers seeking to adopt a Bayesian perspective in statistical practice.
\end{abstract}


\newpage

\section{Introduction}

The axiomatic definition of probability states that it is a function corresponding to a measure from a $\sigma$-algebra of subsets of the sample space to the real interval [0,1]. This concept is the same under both the frequentist and Bayesian approaches to statistics. Nevertheless, it is the interpretation of probability that sets the difference between the two approaches. The subjective interpretation considers probability as an indicator of personal evaluation of an uncertain event. On the other hand, the frequentist interpretation is based on the long run relative frequency of such event \citep{GhoshEtAl2007}. There are situations in which subjective interpretation is more useful. For example, when we are given a probability of rain and have to decide whether to take an umbrella or not. Moreover, when considering events like a particular horse winning a given race, subjective interpretation is the valid one. It is more natural for a person to think about the probability of an event as a quantity describing uncertainty, rather than its relative frequency under the abstract (and quite unnatural) scenario of repeating the experiment under similar conditions. 
 
Several contributions set the philosophical foundations of Bayesian statistics (\citealt{cox1946probability}, \citealt{de1970logical}, and \citealt{savage1972foundations}). In particular, the formulation of statistical inference as a decision problem unifies point estimation, hypothesis testing, and interval estimation. Moreover, this perspective has been further developed and clarified in modern expositions of the Bayesian paradigm (\citealt{BernardoSmith2000} and \citealt{paulino2003estatistica}). This line of work permits the establishment of a formal axiomatic framework for decision making under uncertainty, where Bayes’ rule emerges as the optimal mechanism for updating degrees of belief in the presence of new information. This construction is based on the idea of a decision-making process that avoids certain logical inconsistencies (discussed later).

\section{Historical background, foundations, and scope}

Bayesian inference originated with the 18th-century work of \textit{Thomas Bayes}, whose posthumously published theorem introduced a formal mechanism for updating probabilistic beliefs based on observed evidence \citep{BernardoSmith2000}. \textit{Pierre Simon Laplace} later extended and applied these ideas across scientific domains, introducing the use of uniform prior distributions to reflect ignorance and laying the groundwork for what is now known as objective Bayesian analysis. Although frequentist methods, developed in the 20th century by \textit{Fisher}, \textit{Neyman}, and \textit{Pearson}, became the dominant statistical paradigm by emphasizing long-run frequencies and sampling distributions, Bayesian principles remained influential in decision theory and subjective probability through the contributions of de \textit{Finetti} and \textit{Savage} \citep{GhoshEtAl2007}. The revival of Bayesian methods in the late 20th century was largely driven by advances in computational algorithms, particularly Markov Chain Monte Carlo, which made it feasible to perform inference in complex and high-dimensional models \citep{robert1999monte}. Today, Bayesian inference is a foundational tool in modern statistics and machine learning, valued for its coherent quantification of uncertainty, principled use of prior information, and flexibility in modeling.

While probability and statistics are closely related, they address complementary objectives. Probability theory begins with a specified model and describes the expected behavior of data generated under that model, such as calculating the distribution of outcomes from repeated coin tosses. In contrast, statistics begins with observed data and seeks to infer the underlying structure or parameters responsible for generating those observations. In the Bayesian framework, probability serves both purposes: it models data variability and expresses uncertainty about unknown quantities, enabling inference and prediction within a unified probabilistic system. This contrasts with frequentist approaches, which treat parameters as fixed and base inference on the long-run behavior of estimators. The Bayesian perspective allows for coherent updating of beliefs in light of data, particularly in complex, small-sample, or data-limited settings.

The central goal of the Bayesian paradigm is to update prior beliefs about unknown parameters using observed data, yielding posterior distributions that synthesize both sources of information through Bayes' theorem. These posterior distributions enable direct probabilistic statements about parameters, predictions, and hypotheses without reliance on asymptotic approximations or repeated sampling logic. The Bayesian approach also naturally extends to decision making under uncertainty through the use of utility functions, guiding optimal actions based on posterior beliefs. Its broad applicability includes point estimation, interval estimation, hypothesis testing, model comparison, prediction, and hierarchical modeling, with successful applications across disciplines such as medicine, economics, political science, and artificial intelligence, among many others. The capacity to incorporate prior knowledge and to adapt to complex modeling structures makes Bayesian inference especially powerful in high-dimensional or sparse-data problems.


\section{Bayesian philosophy and mathematical foundations}

Several seminal contributions set the mathematical foundations of the Bayesian approach. Discussing them all is a formidable task beyond the scope of this paper. Thus, for the sake of simplicity, in this section we present a summary of the main ideas and results that establish the formal mathematical support of Bayesian inference, closely following \cite{BernardoSmith2000} and \cite{GhoshEtAl2007}.

The main idea is that Bayes' theorem provides a rational way to update the degree of belief about an uncertain event. The rationality is given by the maximization of a utility function. Bayes' theorem permits handling degrees of belief under new evidence about the uncertain events of interest.

The approach is based on formulating statistical inference problems as decision problems, which constitutes a unifying framework in the sense that point estimation, interval estimation, and hypothesis testing problems may be formally studied under this approximation. A central concept in this construction is \emph{rationality}, which is intended to avoid inconsistencies in a decision-making process under uncertainty.

Let us start by formally defining a decision problem. A decision problem comprises the following collection
\[
\{\varepsilon, C, O, \le\},
\]
where $\varepsilon$ is a set of uncertain events $E_j,\, j\in J$, which contains a certain event denoted by $\Omega$ (the sample space in the usual statistical literature). The philosophical considerations discussed in \citet{BernardoSmith2000} lead to $\varepsilon$ having the structure of a sigma-algebra. Continuing with the description of the elements of a decision problem, $C$ is a set of possible consequences $c_j,\, j\in J$, and $O$ is a set of options or actions corresponding to mappings from a partition of $\Omega$ to the set of consequences. Thus, a typical element of $O$ has the form
\[
\{c_j \mid E_j,\; j\in J\},
\]
meaning that event $E_j$ leads to consequence $c_j$, for $j \in J$. It is worth noticing that $\le$ is a binary relation over $O\times O$ which defines a preference order between options. This relationship induces preferences between consequences and extends to $\varepsilon\times\varepsilon$, where it defines an uncertainty relation between events and allows stating that one event is more likely than another. Moreover, since options or actions are mappings, a composite function notation can be used as follows. The option
\[
a=\{c_1 \mid E\cap G,\; c_2 \mid E^{c}\cap G,\; c_3 \mid G^{c}\},
\]
can be written as
\[
a=\{a_1 \mid G,\; c_3 \mid G^{c}\}, \qquad 
a_1=\{c_1 \mid E,\; c_2 \mid E^{c}\}.
\]
Also, $c=\{c\mid\Omega\}$ is referred to as a \emph{pure} consequence.

When there is new evidence from observed events. For example, once data is observed, preferences can be computed conditioned on the occurrence of such events. In this case, we define the following conditional preference relationships for actions (options), consequences, and uncertainties about events. For any possible event $G$:
\begin{enumerate}[label=\roman*)]
\item $a_1 \le_G a_2 \;\Longleftrightarrow\; \forall a\in O,\; \{a_1 \mid G,\; a\mid G^c\} \le \{a_2 \mid G,\; a\mid G^c\}$.

\item $c_1 \le_G c_2 \;\Longleftrightarrow\; \{c_1\mid\Omega\} \le_G \{c_2\mid\Omega\}$.

\item $E \le_G F \;\Longleftrightarrow\;$ for $c_1 \le_G c_2$,
\[
\{c_2\mid E,\; c_1\mid F\} \le_G \{c_2\mid E,\; c_1\mid F\}.
\]
\end{enumerate}

Now, rational preferences are formally stated in the form of a system of five axioms, which sets the basis for the concept of quantitative coherence. Furthermore, this leads to the fact that the formal definition of the degree of belief about an uncertain event requires it to be defined in terms of a probability measure. So far, it can be noticed that this formal approach to decision making under uncertainty, uses the elements of probability theory. Without going into details, the following is the prescriptive axiomatic system mentioned above. It contains five axioms that can be divided into two stages: the first one sets a coherent qualitative system (three axioms), and the second one introduces quantitative coherence (two axioms).

\subsubsection*{First stage: defining a qualitative coherent system}

\paragraph{Axiom 1. Compatibility of consequences and dichotomized options.}
\begin{enumerate}[label=\roman*)]
\item $\exists\, c_1,c_2$ such that $c_1 < c_2$.
\item For all consequences $c_1,c_2$ and all events $E$ and $F$, either
\[
\{c_2\mid E,\, c_1\mid E^c\} \le \{c_2\mid F,\, c_1\mid F^c\}
\quad\text{or}\quad
\{c_2\mid E,\, c_1\mid E^c\} \ge \{c_2\mid F,\, c_1\mid F^c\}.
\]
\end{enumerate}
If all decisions were equivalent, we would not be facing a decision problem, which is what part (i) states. On the other hand, part (ii) does not state that all options can be compared, it simply ensures that we can make choices between pairs of simple options, which sets the starting point for making rational choices between options.

\paragraph{Axiom 2. Preferences are transitive.}
\begin{enumerate}[label=\roman*)]
\item $a\sim a$.
\item If $a_1\le a_2$ and $a_2\le a_3$, then $a_1\le a_3$.
\end{enumerate}
This axiom induces transitivity of uncertainties, that is, $E\sim E$, and if $E_1\le E_2$ and $E_2\le E_3$, then $E_1\le E_3$. Here the relation $\sim$ is induced by $\le$ as follows:
\[
E_1\sim E_2 \;\Longleftrightarrow\; E_1\le E_2\ \text{and}\ E_2\le E_1.
\]
In this case, we say that $E_1$ is equally likely to $E_2$.

\paragraph{Axiom 3. Consistency of preferences.}
\begin{enumerate}[label=\roman*)]
\item If $c_1\le c_2$, then for all $G\neq\emptyset$, $c_1 \le_G c_2$.
\item If, for some $c_1<c_2$,
\[
\{c_2\mid E,\, c_1\mid E^c\} \le \{c_2\mid F,\, c_1\mid F^c\},
\]
then $E\le F$.
\item If, for some $c$ and $G\neq\emptyset$,
\[
\{a_1\mid G,\, c\mid G^c\} \le \{a_2\mid G,\, c\mid G^c\},
\]
then $a_1 \le_G a_2$.
\end{enumerate}
This axiom presents a formal statement of the idea that preferences between pure consequences should not be affected by new information (part i). Part (ii) states that if we do not prefer $E$ over $F$ for some consequences $c_1<c_2$, then we should have this preference for any pair of consequences, that is, the preference should not depend on particular consequences used to construct an option, but only on the relative likelihood of the events. Lastly, part (iii) states that, if the preference holds for some $c$, then this relation carries over for any action $a$, using the composite-function notation introduced earlier.

\subsubsection*{Second stage: Quantification (quantitative coherence)}

\paragraph{Axiom 4. Existence of standard events.}  
This axiom is composed of five parts describing the existence of a sub-algebra of $\varepsilon$ whose events satisfy properties that make them analogous to measurement units like kg. or cm. For the sake of simplicity, we do not state the axiom formally. However, it is important to briefly discuss the analogy with measuring physical magnitudes. Many measures are defined as any real number. However, the measurement procedure has limited accuracy, resulting in measurements taking values in some subset of the rational numbers. Physical measurement is performed using a standard unit such as cm. There is an implicit continuous scale, but the result is obtained by qualitative pairwise comparison with points in the standard scale. In conclusion, the existence of standard events allows the measurement of preferences and uncertainties, which is precisely the role of the next axiom.

\paragraph{Axiom 5. Precise measurement of preferences and uncertainties.}
\begin{enumerate}[label=\roman*)]
\item If $c_1\le c\le c_2$, there exists a standard event $S$ such that
\[
c \sim \{c_2\mid S,\; c_1\mid S^c\}.
\]
\item For each event $E$ there exists a standard event $S$ such that $E\sim S$.
\end{enumerate}
This axiom introduces a contraction argument somewhat analogous to the ``sandwich theorem''. We can ``sandwich'' a consequence $c$ arbitrarily tightly between options defined in terms of standard events (part i). A similar argument applies for part (ii), but in this case we are dealing with events.

To close this section, we summarize some main results emerging from this axiomatic system:
\begin{itemize}
\item Restricting attention to countably additive probability is the basis for representing beliefs.
\item It turns out that the only decision criterion compatible with the proposed axiom system is maximizing utility (minimizing loss).
\item The uncertainty relation $\le$ (on $\varepsilon\times\varepsilon$) is compatible with a probability measure $P(\cdot)$.
\item The conditional probability measure is the only one compatible with the conditional uncertainty relation $\le_E$.
\item Quantitative coherence implies that degrees of belief must obey the rules of probability measures.
\item Bayes' theorem allows updating degrees of belief when there is new information. Specifically, there exists a loss function $L$ (whose negative can be taken as a utility function) and a prior distribution such that action $a_2$ is at least as good as $a_1$ if, and only if,
\[
\int_\Theta L(\boldsymbol{\theta}, a_2)\,
p(\boldsymbol{\theta}\mid \boldsymbol{y})\, \textsf{d}\boldsymbol{\theta}
\;\le\;
\int_\Theta L(\boldsymbol{\theta}, a_1)\,
p(\boldsymbol{\theta}\mid \boldsymbol{y})\, \textsf{d}\boldsymbol{\theta}.
\]
This equation means that $a_2$ is at least as good as $a_1$ if and only if it has a smaller posterior risk (or greater posterior utility).
\end{itemize}

We want to close this section by emphasizing that, in contrast to other approaches to statistical inference, the Bayesian approach is based on axiomatic foundations leading to a logical framework.

\section{The Bayesian paradigm}

Bayesian inference is rooted in a model-based framework that represents uncertainty about the data-generating process through a probabilistic model. The observed data vector $\boldsymbol{y} = (y_1, \ldots, y_n)$ is assumed to be drawn from a distribution $P$ characterized by a density or probability mass function $p$, which is indexed by a finite-dimensional parameter vector $\boldsymbol{\theta} \in \Theta$. For example, in a normal model with unknown mean and variance, we may write $\boldsymbol{\theta} = (\mu, \sigma^2)$, with $\Theta = \mathbb{R} \times \mathbb{R}^+$. The likelihood function $p(\boldsymbol{y} \mid \boldsymbol{\theta})$ encodes the information the data provide about $\boldsymbol{\theta}$ and is central to all Bayesian updating.

In contrast to frequentist methods, which treat parameters as fixed and base inference on sampling distributions, the Bayesian paradigm treats both data and parameters as random variables. A prior distribution with density or mass function $p(\boldsymbol{\theta})$ is assigned to $\boldsymbol{\theta}$ to formally encode prior uncertainty or external knowledge. This prior may reflect subjective beliefs, expert elicitation, historical information, or intentionally vague assumptions.

Upon observing $\boldsymbol{y}$, \textit{Bayes’ theorem} is used to update beliefs and derive the posterior distribution,
\[
p(\boldsymbol{\theta} \mid \boldsymbol{y}) = \frac{p(\boldsymbol{y} \mid \boldsymbol{\theta}) \, p(\boldsymbol{\theta})}{\int_\Theta p(\boldsymbol{y} \mid \boldsymbol{\theta}) \, p(\boldsymbol{\theta}) \, \textsf{d}\boldsymbol{\theta}},
\]
which combines prior information with the observed data and forms the basis for all Bayesian inference. It is a common practice to denote the likelihood, the prior, and posterior using the symbol $p$, with interpretation depending on context.

Two predictive distributions further enrich the Bayesian framework. The \textit{prior predictive distribution}, given by
\[
p(\boldsymbol{y}) = \int_\Theta p(\boldsymbol{y}, \boldsymbol{\theta}) \, \textsf{d}\boldsymbol{\theta} = \int_\Theta p(\boldsymbol{y} \mid \boldsymbol{\theta}) \, p(\boldsymbol{\theta}) \, \textsf{d}\boldsymbol{\theta},
\]
is the marginal distribution of the data prior to observation and corresponds to the denominator in Bayes’ theorem. The \textit{posterior predictive distribution}, used to make probabilistic statements about future data $\boldsymbol{y}^*$, is given as
\[
p(\boldsymbol{y}^* \mid \boldsymbol{y}) = \int_\Theta p(\boldsymbol{y}^* , \boldsymbol{\theta} \mid \boldsymbol{y}) \, \textsf{d}\boldsymbol{\theta} = \int_\Theta p(\boldsymbol{y}^* \mid \boldsymbol{\theta}) \, p(\boldsymbol{\theta} \mid \boldsymbol{y}) \, \textsf{d}\boldsymbol{\theta},
\]
under the typical assumption that future and observed data are conditionally independent given $\boldsymbol{\theta}$. These predictive distributions are crucial for forecasting, model checking, and decision making under uncertainty. While the models employed are inevitably approximations of reality, the Bayesian approach emphasizes the construction of useful and interpretable representations of uncertainty.

Bayesian models can be broadly classified as parametric or nonparametric, depending on the assumptions made about the structure of the underlying data-generating process. In parametric models, the distribution $p(\boldsymbol{y} \mid \boldsymbol{\theta})$ is fully characterized by a finite-dimensional parameter vector $\boldsymbol{\theta} \in \Theta$. These models are typically more tractable, especially when combined with conjugate priors, and are well suited for problems where strong structural assumptions are reasonable. In contrast, nonparametric Bayesian models place priors on infinite-dimensional spaces, allowing greater flexibility in capturing complex data structures without specifying a fixed parametric form. Examples include Dirichlet process mixtures and Gaussian process models. While computationally more demanding, nonparametric approaches are valuable when the underlying distribution is unknown or difficult to specify, enabling more adaptive inference while still adhering to Bayesian principles.

\subsection{Example: Normal-Normal model}

Consider a case where the data $\boldsymbol{y} = (y_1, \ldots, y_n)$ consist of an independent and identically distributed sample from a normal distribution with known variance and unknown mean. Specifically, assume that $y_i \mid \theta \overset{\text{i.i.d.}}{\sim} \textsf{N}(\theta, 1)$ for $i = 1, \ldots, n$, and that the parameter $\theta$ has a Gaussian prior $\theta \sim \textsf{N}(\mu, \tau^2)$, where $\mu$ and $\tau^2$ are known hyperparameters representing prior beliefs about the location and variability of $\theta$.

In this case, the likelihood function is given by
$$
p(\boldsymbol{y} \mid \theta) = \prod_{i=1}^n \frac{1}{\sqrt{2\pi}} \exp\left\{-\frac{1}{2}(y_i - \theta)^2\right\}
\propto \exp\left\{-\frac{1}{2} \sum_{i=1}^n (y_i - \theta)^2\right\} 
\propto \exp\left\{-\frac{n}{2}(\bar{y} - \theta)^2\right\},
$$
since $\sum_{i=1}^n (y_i - \theta)^2 = \sum_{i=1}^n (y_i - \bar{y})^2 + n(\bar{y} - \theta)^2$, where $\bar{y} = \frac{1}{n} \sum_{i=1}^n y_i$ denotes the sample mean. The prior density corresponding to $\theta \sim \textsf{N}(\mu, \tau^2)$ is
\[
p(\theta) \propto \exp\left\{-\frac{1}{2\tau^2}(\theta - \mu)^2\right\}.
\]
Combining the likelihood and the prior yields the unnormalized posterior:
$$
p(\theta \mid \boldsymbol{y}) \propto \exp\left\{-\frac{n}{2}(\theta - \bar{y})^2 - \frac{1}{2\tau^2}(\theta - \mu)^2\right\} 
\propto \exp\left\{ -\frac{1}{2} \left[\left(\frac{1}{\tau^2} + n\right)\theta^2 - 2\left(\frac{\mu}{\tau^2} + n\bar{y}\right)\theta \right] \right\},
$$
which corresponds to the kernel of a normal distribution with posterior mean and variance given by
\[
\theta_n = \frac{\frac{1}{\tau^2}\mu + n\bar{y}}{\frac{1}{\tau^2} + n} 
\qquad \text{and} \qquad
\tau_n^2 = \frac{1}{\frac{1}{\tau^2} + n}.
\]
Therefore, the posterior distribution is $\theta \mid \boldsymbol{y} \sim \textsf{N}(\theta_n, \tau_n^2)$.

To make predictions for a new observation $y^*$, we compute the posterior predictive distribution
\[
p(y^* \mid \boldsymbol{y}) = \int_{-\infty}^\infty \textsf{N}(y^* \mid \theta, 1) \, \textsf{N}(\theta \mid \theta_n, \tau_n^2) \, \textsf{d}\theta,
\]
which corresponds to the convolution of two normal densities. The result is that $y^* \mid \boldsymbol{y} \sim \textsf{N}(\theta_n, \tau_n^2 + 1)$. This distribution reflects both the inherent variability in the data and the residual uncertainty in the parameter $\theta$. As the sample size $n$ increases, the posterior variance $\tau_n^2$ tends to zero, and the predictive variance converges to $1$, the variance of the data-generating distribution. \hfill$\square$

\section{Conjugate models}

A family of prior distributions $\mathcal{C}$ is said to be \textit{conjugate} to a likelihood function $p(\boldsymbol{y} \mid \theta)$ if, for any prior $p(\theta)$ in $\mathcal{C}$, the resulting posterior distribution $p(\theta \mid \boldsymbol{y})$ also belongs to $\mathcal{C}$. That is, if $p(\theta \mid \boldsymbol{y}) \in \mathcal{C}$, then the family is closed under Bayesian updating with respect to the likelihood $p(\boldsymbol{y} \mid \theta)$. Conjugate priors are particularly useful in analytical settings, as they yield posterior distributions with the same functional form as the prior, facilitating closed-form expressions for posterior quantities such as means, variances, and predictive distributions.

\subsection{Example: Binomial-Beta model}

Consider a binomial likelihood where $y \mid \theta \sim \textsf{Binomial}(n, \theta)$, with $n$ known and $\theta$ unknown. The likelihood function is
\[
p(y \mid \theta) = \binom{n}{y} \theta^y (1 - \theta)^{n - y}, \quad \theta \in [0,1], \quad y \in \{0,1,\ldots,n\}.
\]
We assign a prior from the Beta family,
\[
p(\theta) \propto \theta^{a - 1}(1 - \theta)^{b - 1}, \quad a > 0, \quad b > 0.
\]
The unnormalized posterior is the product of the likelihood and the prior:
\[
p(\theta \mid y) \propto \theta^y (1 - \theta)^{n - y} \cdot \theta^{a - 1}(1 - \theta)^{b - 1} = \theta^{a + y - 1}(1 - \theta)^{b + n - y - 1}.
\]
Recognizing this as the kernel of a Beta distribution, the normalized posterior is $\theta \mid y \sim \textsf{Beta}(a + y, \, b + n - y)$. Thus, the Beta prior is conjugate to the Binomial likelihood, and posterior inference remains within the Beta family, with parameters updated by combining prior information and observed data. \hfill$\square$

\subsection{Example: Poisson-Gamma model}

Consider a Poisson likelihood where $y_1, \ldots, y_n$ are independent observations such that $y_i \mid \theta \overset{\text{i.i.d.}}{\sim} \textsf{Poisson}(\theta)$, with $\theta > 0$ unknown. The likelihood function is
\[
p(\boldsymbol{y} \mid \theta) = \prod_{i=1}^n \frac{\theta^{y_i} e^{-\theta}}{y_i!} \propto \theta^{\sum_{i=1}^n y_i} e^{-n\theta}, \quad \theta > 0, \quad y_i \in \{0,1,2,\ldots\}.
\]
We assign a prior from the Gamma family,
\[
p(\theta) \propto \theta^{a - 1} e^{-b\theta}, \quad a > 0, \quad b > 0.
\]
The unnormalized posterior is the product of the likelihood and the prior:
\[
p(\theta \mid \boldsymbol{y}) \propto \theta^{\sum y_i} e^{-n\theta} \cdot \theta^{a - 1} e^{-b\theta} = \theta^{a + \sum y_i - 1} e^{-(b + n)\theta}.
\]
Recognizing this as the kernel of a Gamma distribution, the normalized posterior is $\theta \mid \boldsymbol{y} \sim \textsf{Gamma}(a + \textstyle\sum_{i=1}^n y_i, \, b + n)$. Thus, the Gamma prior is conjugate to the Poisson likelihood, and posterior inference remains within the Gamma family, with updated parameters that reflect both prior beliefs and the cumulative information from the observed counts. \hfill$\square$

\subsection{Example: Exponential family}

A particularly elegant property of the natural exponential family is its closure under conjugate priors. In the scalar case, the likelihood for data $\boldsymbol{y}$ can be written as
\[
p(\boldsymbol{y} \mid \theta) = \exp\left\{ \theta \, t(\boldsymbol{y}) + g(\theta) + h(\boldsymbol{y}) \right\},
\]
where $t(\boldsymbol{y})$ is a sufficient statistic, $g(\theta)$ captures the dependence on the natural parameter $\theta$, and $h(\boldsymbol{y})$ depends only on the data. A conjugate prior for $\theta$ takes the form
\[
p(\theta) = \exp\left\{ \theta \mu + \nu g(\theta) + c(\mu, \nu) \right\},
\]
where $\mu$ is the prior location parameter, $\nu > 0$ plays the role of prior precision (equivalent sample size), and $c(\mu, \nu)$ ensures proper normalization. This prior also belongs to the exponential family and yields a posterior distribution of the same form:
\[
p(\theta \mid \boldsymbol{y}) = \exp\left\{ \theta (\mu + t(\boldsymbol{y})) + (\nu + 1) g(\theta) + c(\mu + t(\boldsymbol{y}), \nu + 1) \right\}.
\]
This result demonstrates that the exponential family is closed under Bayesian updating with conjugate priors, offering both analytical tractability and interpretability. Later, we present a formal result characterizing conjugate priors in natural exponential families.\hfill$\square$

\subsection{Example: Linear Bayes}

In both the Binomial-Beta and Poisson-Gamma examples, the posterior mean of $\theta$ can be expressed as a weighted average of the prior mean and the maximum likelihood estimator (MLE). Specifically, in the Binomial-Beta case, the posterior mean is
\[
\textsf{E}[\theta \mid y] = \frac{a + y}{a + b + n} = \left( \frac{a + b}{a + b + n} \right) \cdot \frac{a}{a + b} + \left( \frac{n}{a + b + n} \right) \cdot \frac{y}{n},
\]
which is a convex combination of the prior mean $a / (a + b)$ and the MLE $y/n$. Similarly, in the Poisson-Gamma model, the posterior mean is
\[
\textsf{E}[\theta \mid \boldsymbol{y}] = \frac{a + \sum y_i}{b + n} = \left( \frac{b}{b + n} \right) \cdot \frac{a}{b} + \left( \frac{n}{b + n} \right) \cdot \bar{y},
\]
where $a / b$ is the prior mean and $\bar{y} = \sum y_i / n$ is the sample mean (MLE). An analogous expression holds in the Gaussian case discussed earlier. These expressions illustrate a key feature of conjugate Bayesian updating: the posterior mean results from a precision-weighted average of the prior mean and the MLE, a property often referred to as \textit{linear Bayes}. \hfill$\square$

\subsection{Example: Mixtures of conjugate priors}

An important and flexible extension of the conjugate framework arises when employing discrete mixtures of conjugate priors. Let \( p(\boldsymbol{\theta}) \) be a prior defined as a finite mixture of \( K \) conjugate components:
\[
p(\boldsymbol{\theta}) = \sum_{k=1}^K \omega_k \, p_k(\boldsymbol{\theta}),
\]
where \( \omega_k \in (0,1) \), with \( \sum_{k=1}^K \omega_k = 1 \), and each component \( p_k(\boldsymbol{\theta}) \) is conjugate to the likelihood \( p(\boldsymbol{y} \mid \boldsymbol{\theta}) \). Under this specification, the posterior is also a finite mixture of conjugate posteriors:
\[
p(\boldsymbol{\theta} \mid \boldsymbol{y}) = \sum_{k=1}^K \omega_k^* \, p_k^*(\boldsymbol{\theta}),
\]
where each component \( p_k^*(\boldsymbol{\theta}) \) is obtained by updating \( p_k(\boldsymbol{\theta}) \) via Bayes' theorem:
\[
p_k^*(\boldsymbol{\theta}) = \frac{p(\boldsymbol{y} \mid \boldsymbol{\theta}) \, p_k(\boldsymbol{\theta})}{\int_\Theta p(\boldsymbol{y} \mid \boldsymbol{\theta}) \, p_k(\boldsymbol{\theta}) \, \textsf{d}\boldsymbol{\theta}},
\]
and the updated mixture weights \( \omega_k^* \) are given by
\[
\omega_k^* = \frac{\omega_k \, \int_\Theta p(\boldsymbol{y} \mid \boldsymbol{\theta}) \, p_k(\boldsymbol{\theta}) \, \textsf{d}\boldsymbol{\theta}}{\sum_{\ell=1}^K \omega_\ell \, \int_\Theta p(\boldsymbol{y} \mid \boldsymbol{\theta}) \, p_\ell(\boldsymbol{\theta}) \, \textsf{d}\boldsymbol{\theta}}.
\]
This formulation allows for the representation of more flexible prior beliefs, such as multimodality, asymmetry, or heavy tails, while retaining the computational and analytical tractability of conjugate updating. Each component is updated independently according to Bayes' rule, and the overall posterior remains a weighted sum of the resulting component posteriors. \hfill$\square$

\subsection{Example: Conjugacy beyond the exponential family}

While many standard examples of conjugate priors arise naturally within the exponential family framework, conjugacy is not limited to this class. A notable counterexample is the case where the data are independently and identically distributed from a uniform distribution on the interval \([0, \theta]\), and the prior for \( \theta \) is a Pareto distribution.

Specifically, suppose \( y_i \mid \theta \overset{\text{i.i.d.}}{\sim} \textsf{Uniform}[0, \theta] \), for \( i = 1, \ldots, n \), and assume a prior \( \theta \sim \textsf{Pareto}(a, b) \), with density
\[
p(\theta) \propto \frac{1}{\theta^{a + 1}} \, \mathbb{I}(\theta \geq b),
\]
where $\mathbb{I}(\cdot)$ is the indicator function. The likelihood for the sample \( \boldsymbol{y} = (y_1, \ldots, y_n) \) is given by
\[
p(\boldsymbol{y} \mid \theta) = \frac{1}{\theta^n} \, \mathbb{I}(\theta \geq y_{(n)}),
\]
where \( y_{(n)} = \max \{ y_1, \ldots, y_n \} \) is the sample maximum. Combining the prior and the likelihood yields the unnormalized posterior:
\[
p(\theta \mid \boldsymbol{y}) \propto \frac{1}{\theta^n} \cdot \frac{1}{\theta^{a + 1}} \cdot \mathbb{I}(\theta \geq b) \cdot \mathbb{I}(\theta \geq y_{(n)}) = \frac{1}{\theta^{a + n + 1}} \, \mathbb{I}\left( \theta \geq \max( b, y_{(n)} ) \right).
\]
This is the kernel of a Pareto distribution with updated parameters:
\[
\theta \mid \boldsymbol{y} \sim \textsf{Pareto}\big(a + n, \max( b, y_{(n)} )\big).
\]
Thus, although this model does not belong to the exponential family, the Pareto prior remains conjugate to the uniform likelihood. This provides a clear example of conjugacy arising outside the exponential family context. \hfill$\square$

\subsection{Example: Beyond conjugate models}

When conjugacy is not available, the Bayesian updating process remains conceptually the same, but analytical progress is often limited. Nevertheless, in some cases, non-conjugate priors can still lead to tractable posterior forms. A well-known example involves the use of a double exponential (Laplace) prior for the location parameter \(\theta\) in a Gaussian likelihood.

Specifically, suppose \( y_i \mid \theta \overset{\text{i.i.d.}}{\sim} \textsf{N}(\theta, 1) \), for \( i = 1,\ldots,n \), and assume a prior \( \theta \sim \textsf{Laplace}(\mu, \tau) \) with density
\[
p(\theta) = \frac{1}{2\tau} \exp\left\{ -\frac{1}{\tau} |\theta - \mu| \right\},
\]
where \( \mu \in \mathbb{R} \) is the prior location and \( \tau > 0 \) is the scale parameter. In this setting, the posterior distribution takes the form of a mixture of truncated normal distributions:
\[
p(\theta \mid \boldsymbol{y}) = \frac{a}{a + b} \, \textsf{N}_{(-\infty, \mu )} \left( \lambda_1, 1/n \right) + \frac{b}{a + b} \, \textsf{N}_{(\mu, \infty )} \left( \lambda_2, 1/n \right),
\]
where the component means are given by \( \lambda_1 = \lambda_1(n, \bar{y}) = \bar{y} + \frac{1}{n\tau} \) and \( \lambda_2 = \lambda_2(n, \bar{y}) = \bar{y} - \frac{1}{n\tau} \), with \( \bar{y} = \frac{1}{n} \sum_{i=1}^n y_i \) denoting the sample mean.

The weights \( a = a(n, \bar{y}) \) and \( b = b(n, \bar{y}) \) serve as normalizing constants for the truncated components. In contrast to the linear form of the posterior mean under a normal prior, the posterior arising from a Laplace prior is highly non-linear in both the data and the prior parameters. A particularly important property of the Laplace prior is its bounded influence: even as the prior center \( \mu \) tends to \( \pm \infty \), the posterior mean remains within a fixed distance of the sample mean. This robustness makes it especially attractive in settings where prior misspecification is a concern.

\subsection{Characterizing conjugate priors in natural exponential families: The Diaconis-Ylvisaker prior}

For natural exponential families, there are results that permit finding non-subjective conjugate priors based on the idea of imitating the likelihood, in the sense that the kernel of the prior density mimics the likelihood. Moreover, these priors enjoy a posterior linearity property. In the following, we discuss results due to \citet{diaconis1979conjugate} that formalize and unify properties discussed in sections 5.3 and 5.4.

Consider the natural exponential family (NEF) of probability distributions whose probability density function can be written in the form
\begin{align*}
  p(\boldsymbol{y} \mid \boldsymbol{\theta})
  =
  h(\boldsymbol{y})
  \exp\Big\{
    \boldsymbol{\theta}^{\top}\boldsymbol{y} - \psi(\boldsymbol{\theta})
  \Big\},
  \qquad \boldsymbol{\theta} \in \boldsymbol{\Theta} \subseteq \mathbb{R}^{p},
\end{align*}
where $h(\cdot)$ is a known function depending on $\boldsymbol{y}$, and $\psi(\cdot)$ is a known function depending on the model parameters $\boldsymbol{\theta}$ that defines the cumulant generating function. It follows that the usual regularity condition of differentiation and integration exchange holds for NEF, thus, it can be shown that the gradient vector of $\psi(\boldsymbol{\theta})$ is the mean parameter of the distribution, that is,
\begin{align*}
  \nabla \psi(\boldsymbol{\theta})
  =
  \textsf{E}(\boldsymbol{Y} \mid \boldsymbol{\theta}).
\end{align*}

Consider the family of prior densities $p(\boldsymbol{\theta})$, defined by the collection of integrable functions
\begin{align*}
  p(\boldsymbol{\theta} \mid n_{0}, \boldsymbol{x}_{0})
  \propto
  \exp\Big\{
    n_{0}\big(
      \boldsymbol{\theta}^{\top}\boldsymbol{x}_{0} - \psi(\boldsymbol{\theta})
    \big)
  \Big\},
\end{align*}
thus, $p(\boldsymbol{\theta} \mid n_{0}, \boldsymbol{x}_{0})$ is an unnormalized density. The natural exponential family of probability distributions is defined through a sigma-finite measure on the Borel sigma-algebra over $\mathbb{R}^{p}$, noted as $\nu$. Besides, let $\mathcal{X}$ be the interior of the convex hull of the support set of $\nu$. Under this setting, \citet{diaconis1979conjugate} showed that $p(\boldsymbol{\theta} \mid n_{0}, \boldsymbol{x}_{0})$ can be normalized, that is, it is integrable if $\boldsymbol{x}_{0} \in \mathcal{X}$ and $n_{0} > 0$. Moreover, if $p(\boldsymbol{\theta} \mid n_{0}, \boldsymbol{x}_{0})$ is integrable and $\boldsymbol{\Theta}=\mathbb{R}^{p}$ then $n_{0} > 0$, while if $p(\boldsymbol{\theta} \mid n_{0}, \boldsymbol{x}_{0})$ is integrable and $n_{0} > 0$ then $\boldsymbol{x}_{0} \in \mathcal{X}$. In addition, it follows that if $\boldsymbol{\theta}$ has distribution with density $p(\boldsymbol{\theta} \mid n_{0}, \boldsymbol{x}_{0})$, then the prior expected value of the mean parameter is
\begin{align*}
  \textsf{E}\big(\nabla \psi(\boldsymbol{\theta})\big)
  =
  \boldsymbol{x}_{0}.
\end{align*}

Now we summarize the main result in \citet{diaconis1979conjugate}, which characterizes conjugate priors for NEF. For a Bayesian model with a likelihood pertaining to the NEF, data $\boldsymbol{y}_{1},\ldots,\boldsymbol{y}_{n}$, and a prior $p(\boldsymbol{\theta} \mid n_{0}, \boldsymbol{x}_{0})$ (which implies that the conditions on the hyperparameters for the prior to be proper are satisfied), it follows that:
\begin{enumerate}[label=\roman*)]
  \item The posterior is of the form $p(\boldsymbol{\theta} \mid n_{0}+n, \boldsymbol{x}_{n})$, where
  \begin{align*}
    \bar{\boldsymbol{y}} = \frac{1}{n}\sum_{i=1}^{n}\boldsymbol{y}_{i},
    \qquad
    \boldsymbol{x}_{n}
    =
    \frac{n_{0}\boldsymbol{x}_{0} + n\bar{\boldsymbol{y}}}{n_{0}+n}.
  \end{align*}
  This shows conjugacy.

  \item The posterior expectation of the mean parameter satisfies
  \begin{align*}
    \textsf{E}\big(\nabla \psi(\boldsymbol{\theta}) \mid \boldsymbol{y}_{1},\ldots,\boldsymbol{y}_{n}\big)
    =
    \frac{n_{0}}{n_{0}+n}\,\boldsymbol{x}_{0}
    +
    \frac{n}{n_{0}+n}\,\bar{\boldsymbol{y}}.
  \end{align*}
  Notice that this is a convex combination of the prior expectation of the mean parameter and the sample mean. This property corresponds to linearity of the posterior expectation of the mean parameter and it is known as posterior linearity. An interpretation of the Diaconis--Ylvisaker prior (DY prior) is that $\boldsymbol{x}_{0}$ is the mean of $n_{0}$ observations from the past.

  \item If $p(\boldsymbol{\theta})$ is a prior distribution that is not concentrated at a single point of $\boldsymbol{\Theta}$ (which is open in $\mathbb{R}^p$), $\boldsymbol{y}$ is a realization of a sample of size $1$, and
  \begin{align*}
    \textsf{E}\big(\nabla \psi(\boldsymbol{\theta}) \mid \boldsymbol{y}\big)\
    =
    a\,\boldsymbol{y} + \boldsymbol{b},
  \end{align*}
  for some constant $a$ and constant vector $\boldsymbol{b}$, then $a\neq0$, and the prior density has the form
  \begin{align*}
    p(\boldsymbol{\theta})
    =
    c
    \exp\Big\{
      a^{-1}\big(
        \boldsymbol{\theta}^{\top}\boldsymbol{b} - (1-a)\psi(\boldsymbol{\theta})
      \big)
    \Big\},
  \end{align*}
  where $c$ is the inverse normalizing constant. Since $p(\boldsymbol{\theta})$ has the form of the DY prior, this result characterizes this family of conjugate priors.
\end{enumerate}

Due to the interpretability and conjugacy of this family of priors, it has been used in different inference problems. One of contemporary relevance is graphical models, in particular, Gaussian graphical models, where the DY class of prior densities is known as the $G$-Wishart \citep{roverato2002hyper}, which is used to estimate a concentration matrix with structural zeros dictated by an undirected graph. This family has been widely used to infer concentration matrices in high-dimensional settings.

\section{Bayesian asymptotics}

If the observations \( y_1, \ldots, y_n \) are independent and identically distributed conditional on \( \boldsymbol{\theta} \), the posterior distribution often admits a tractable asymptotic approximation. Under standard regularity conditions, such as differentiability of the log-likelihood, existence and positive definiteness of the observed Fisher information matrix, and posterior consistency, the posterior distribution can be approximated by a multivariate normal distribution as \( n \to \infty \):
\[
\boldsymbol{\theta} \mid \boldsymbol{y} \approx \textsf{N}\big( \hat{\boldsymbol{\theta}}, \, \mathbf{I}^{-1} \big),
\]
where \( \hat{\boldsymbol{\theta}} = \hat{\boldsymbol{\theta}}(\boldsymbol{y}) \) is the maximum likelihood estimator (MLE) of \( \boldsymbol{\theta} \), and \( \mathbf{I} = \mathbf{I}(\hat{\boldsymbol{\theta}}) \) is the observed information matrix, defined by
\[
\mathbf{I}(\hat{\boldsymbol{\theta}}) = -\left. \frac{\partial^2 }{\partial \boldsymbol{\theta} \, \partial \boldsymbol{\theta}^\top} \log p(\boldsymbol{y} \mid \boldsymbol{\theta}) \right|_{\boldsymbol{\theta} = \hat{\boldsymbol{\theta}}}.
\]

This result, often referred to as the \textit{Bayesian Central Limit Theorem}, implies that for large sample sizes, the posterior distribution becomes approximately Gaussian, centered at the MLE, with covariance determined by the curvature of the log-likelihood function. Importantly, this asymptotic approximation is typically independent of the prior, provided the prior is continuous and strictly positive in a neighborhood of the ``true'' parameter value, and its influence diminishes as the sample size increases. Such ``true'' parameter value may be formally envisaged as a form of strong law limit of observables \citep{BernardoSmith2000}. 

Formally, posterior normality may be established under four regularity conditions, relevant contributions to this result are due to works by Bernstein and von-Mises, which is why some statisticians refer to it as the Bernstein-von-Mises theorem \citep{sarkar2023high}. Recent expansions of this result can be found in areas like high-dimensional covariance estimation \citep{sarkar2023high}, and linear regression models with hierarchical g-priors and nonlocal priors  \citep{fang2024bernstein}. Next, we present a simplified version for the case of a real-valued parameter presented in \cite{GhoshEtAl2007}.

\textbf{Theorem.} Let \(X_1,\ldots,X_n\) be an i.i.d.\ sample from the distribution \(P_\theta\), which admits density \(p(x \mid \theta)\), with parameter space $\Theta$ and support $\mathcal{X}$. In the following, the probability statements are made under $\theta_0$, which is regarded as the ``true parameter value''. Consider the following regularity conditions:

\begin{enumerate}[label=\roman*)]
\item The set \{$x:p(x\mid\theta)>0\}$ is the same for all $\theta\in\Theta$. It means that the support set does not depend on $\theta$.

\item The function $l(x\mid\theta) =\log p(x\mid\theta)$ is thrice differentiable with respect to $\theta$ in a neighborhood of $\theta_0$. Moreover, the expected values of the first and second derivatives of $\log p(x\mid\theta)$ are finite and there exists a function $M(\cdot)$ with $\textsf{E}_{\theta_0}\big(M(X)\big)<\infty$ such that:
\[
\sup_{\theta\in(\theta_0-\delta,\theta_0+\delta)}|l^{(3)}(x\mid\theta)|\leq M(x),
\]
where $l^{(3)}$ denotes the third derivative of function $l$.

\item We can interchange the order of integration and differentiation. 

\item Let $L_n(\cdot)$ be the log-likelihood function. For any $\delta>0$, with $P_{\theta_0}$-probability one, it follows that:
\[
\sup_{|\theta-\theta_0|>\delta}(1/n)(L_n(\theta)-L_n(\theta_0)\leq -\epsilon,
\]
for some $\epsilon>0$ and $n$ large enough. 
\end{enumerate}

Besides, let $\hat{\theta}_n$ be a strongly consistent solution to the likelihood equations based on a sample $\boldsymbol{x}=(x_1,\ldots,x_n)$. Then, for any prior density $p(\theta)$ that is continuous and positive at $\theta_0$ it follows that
\[
\lim_{n\to\infty}\int_{\mathbb{R}}\Bigl|\pi_n(t\mid \boldsymbol{x})-\frac{\sqrt{I(\theta_0)}}{\sqrt{2\pi}}
\exp{\Big(-\tfrac{1}{2}\,t^2 I(\theta_0)\Big)}\Bigr| \,\textsf{d}t=0,
\]
where $\pi_n(t\mid \boldsymbol{x})$ is the posterior density of $t=\sqrt{\theta-\hat{\theta}_n}$ and $I(\theta_0)$ corresponds to the Fisher information per unit observation \citep{GhoshEtAl2007}. Furthermore, the result also holds when replacing $I(\theta_0)$ by $-(1/n)L_{n}^{(2)}(\hat{\theta}_n)$ where $L_{n}^{(2)}$ is the second derivative of the log-likelihood evaluated at 

This result indicates that the posterior distribution of $t$ converges in total variation to a Gaussian distribution with mean zero and variance ${I(\theta_0)}^{-1}$, and reconciles frequentist and Bayesian approaches for the large sample scenario because it follows that, for $n$ large enough, the maximum likelihood and Bayes point estimators, as well as the confidence and credible intervals, agree. 

\subsection{Laplace expansions}

Consider a sequence \( y_1, \ldots, y_n \) of independent and identically distributed observations, conditional on a continuous parameter \( \theta \in \mathbb{R} \). Let \( L(\theta) = p(\boldsymbol{y} \mid \theta) \) denote the likelihood function based on the full sample \( \boldsymbol{y} = (y_1, \ldots, y_n) \). 

A classical asymptotic approach to approximating the posterior distribution involves performing a second-order Taylor expansion of the log-likelihood function around the MLE \( \hat{\theta} \). Specifically, we expand the log-likelihood around \( \hat{\theta} \) as
\[
\log L(\theta) \approx \log L(\hat{\theta}) - \frac{1}{2} I(\hat{\theta}) (\theta - \hat{\theta})^2,
\]
where \( I = I(\hat{\theta}) = -\left. \frac{\partial^2}{\partial \theta^2} \log L(\theta) \right|_{\theta = \hat{\theta}} \) is the observed Fisher information evaluated at the MLE. Exponentiating both sides yields a normal approximation to the likelihood:
\[
L(\theta) \approx L(\hat{\theta}) \, \exp\left\{ - \frac{1}{2} I(\hat{\theta}) (\theta - \hat{\theta})^2 \right\}.
\]

The prior \( p(\theta) \) can also be locally approximated via a Taylor expansion around \( \hat{\theta} \). However, under standard regularity conditions, the likelihood becomes increasingly concentrated as \( n \to \infty \), and the influence of the prior on the posterior diminishes. As a result, the posterior distribution is once again approximately normal, centered at the MLE and with variance given by the inverse of the observed Fisher information:
\[
\theta \mid \boldsymbol{y} \approx \textsf{N}\big( \hat{\theta}, \, I^{-1} \big).
\]
The extension to the multivariate case, where \( \boldsymbol{\theta} \in \mathbb{R}^d \), is entirely analogous, with \( I(\hat{\theta}) \) replaced by the observed information matrix \( \mathbf{I}(\hat{\boldsymbol{\theta}}) \).

This approximation is known as the \textit{Laplace approximation}. It differs subtly from the earlier result referred to as the Bayesian Central Limit Theorem, which also yields a normal approximation but arises from more general asymptotic theory involving posterior consistency and convergence in distribution to a Gaussian centered at the true parameter value. In contrast, the Laplace approximation is a purely analytical technique, obtained by locally expanding the posterior density around its mode and is applicable even in models that may not fully satisfy the regularity conditions required for asymptotic convergence. 

An important consequence of the Laplace approximation is that, for large sample sizes, Bayesian and frequentist methods for point estimation and interval estimation tend to yield similar results \citep{GhoshEtAl2007}. However, this asymptotic agreement does not generally extend to hypothesis testing, where prior assumptions continue to play a fundamental influence. As the sample size increases, the likelihood becomes increasingly concentrated around the MLE, causing the impact of the prior to diminish, provided that the prior is continuous and strictly positive in a neighborhood of the true parameter value. Finally, the quality of the normal approximation can be sensitive to the choice of parameterization. In some cases, reparameterizing the model can significantly improve or worsen the accuracy of the Laplace approximation.

The Laplace estimation may be used to find an analytical approximation to integrals of the form
\[
\int_\Theta g(\boldsymbol{\theta})\,p(\boldsymbol{x}\mid\boldsymbol{\theta}) \,\textsf{d}\boldsymbol{\theta}
\]
which appears when computing posterior moments of functions of $\boldsymbol{\theta}$. Consider an integral of the form
\[
I=\int_\Theta q(\boldsymbol{\theta})\,\exp\big(n\,h(\boldsymbol{\theta})\big)\,\textsf{d}\boldsymbol{\theta},
\]
where $q(\cdot)$ and $h(\cdot)$ are smooth functions and $h(\cdot)$ has a unique ``sharp'' maximum at $\hat{\boldsymbol\theta}$. This assumption implies that most of the contribution to the integral comes from a neighborhood around $\hat{\boldsymbol\theta}$. The Laplace approximation to integrals of this form uses the ideas described previously. It uses a second-order Taylor series expansion around $\hat{\boldsymbol\theta}$. The result of this technique is an analytical approximation to $I$ of order $O(n^{-1})$ which has the form
\[
I\approx \frac{(2\pi)^p\,\exp\!{\big(n\,h(\hat{\boldsymbol{\theta}}) \big)} \, q(\hat{\boldsymbol{\theta}})}{n^{p/2}\,
\det\big(\mathbf{H}_h(\hat{\boldsymbol{\theta}})\big)^{1/2}},
\]
where $\mathbf{H}_h(\hat{\boldsymbol{\theta}})$ is the Hessian matrix of $h$ evaluated at $\hat{\boldsymbol\theta}$. This formula can be used out of the field of statistics. For instance, it can be used to derive the well-known Stirling's approximation to $n!$. Now, in Bayesian statistics, $n\,h(\hat{\boldsymbol{\theta}})$ is usually taken to be the log-likelihood function or the logarithm of the un-normalized posterior density. It has been used to approximate Bayes factors and fractional Bayes factors, and to derive the so-called Bayesian information criterion (BIC) for model comparison, which corresponds to the logarithm of the integrated or marginal likelihood function. Applications in statistical-genomics may be found in \cite{martinez2017jointb}. Besides, when approximating the posterior expectation of a positive real-valued function of $\boldsymbol{\theta}$, there is a refinement derived by \cite{tierney1986accurate} and Kass et al. (1988) (cited by \citealt{GhoshEtAl2007}), hence it has been termed Tierney-Kadane-Kass refinement by \cite{GhoshEtAl2007} and it yields an $O(n^{-2})$ approximation. 

\subsection{Asymptotics in a discrete case}

Consider a sequence \( y_1, \ldots, y_n \) of independent and identically distributed observations, conditional on a parameter \( \theta \in \mathbb{R} \) that takes values in a discrete parameter space \( \Theta = \{ \theta_1, \ldots, \theta_K \} \). Suppose the true data-generating distribution corresponds to some \( \theta^* \in \Theta^* \), which may or may not belong to \( \Theta \). In other words, the model may be misspecified in the sense that the true distribution is not necessarily included in the assumed model class.

Let \( \tilde{k} \) denote the index of the parameter value \( \theta_{\tilde{k}} \in \Theta \) that is closest to the true data-generating distribution in the sense of minimizing the Kullback–Leibler (KL) divergence. More precisely,
\[
\tilde{k} = \arg \min_k \int_\mathcal{Y} p(y \mid \theta^*) \, \log  \frac{p(y \mid \theta^*)}{p(y \mid \theta_k)}  \, \textsf{d}y,
\]
so that \( \theta_{\tilde{k}} \) is the KL projection of the true model onto the discrete parameter space \( \Theta \). Under standard regularity conditions, such as that the KL divergence is finite for all \( \theta_k \in \Theta \), uniquely minimized at \( \theta_{\tilde{k}} \), and that the prior assigns positive mass to each \( \theta_k \), the posterior distribution concentrates on \( \theta_{\tilde{k}} \) as the sample size grows. That is,
\[
\lim_{n \to \infty} p(\theta_{\tilde{k}} \mid \boldsymbol{y}) = 1, \qquad \text{and} \qquad \lim_{n \to \infty} p(\theta_k \mid \boldsymbol{y}) = 0 \quad \text{for all } k \neq \tilde{k}.
\]
This result provides a strong consistency guarantee in the context of discrete model selection. Even when the model is misspecified, i.e., when the true distribution does not lie in the parameter space \( \Theta \), the posterior asymptotically concentrates on the model that is closest to the truth in KL divergence.

\section{Bayesian point estimation}

Bayesian point estimation refers to the process of selecting a single value of the parameter \( \theta \) as a representative summary of the posterior distribution \( p(\theta \mid \boldsymbol{y}) \). From a formal decision-theoretic perspective, the choice of point estimate should be guided by the principle of minimizing expected loss. Given a loss function \( L(a, \theta) \), where \( a \) is the action (in this case, a candidate estimate of \( \theta \)), the Bayes estimator is defined as the value of \( a \) that minimizes the posterior expected loss:
\[
\hat{\theta} = \arg\min_a \textsf{E}(L(a, \theta)\mid \boldsymbol{y}) = \arg\min_a \int_\Theta L(a, \theta) \, p(\theta \mid \boldsymbol{y}) \, \textsf{d}\theta.
\]
In principle, the loss function \( L(a, \theta) \) should reflect the specific goals, trade-offs, and consequences relevant to the application at hand. However, in practice, eliciting a meaningful and accurate utility or loss function from a domain expert or decision-maker is often challenging. It may be difficult to quantify preferences, capture all relevant costs, or justify the choice of one loss function over another.

As a result, most Bayesian analyses adopt standard loss functions that lead to widely used point summaries of the posterior distribution, typically interpreted as centrality measures. For example, squared error loss yields the posterior mean, absolute error loss leads to the posterior median, and certain discontinuous losses can justify the use of the posterior mode. These default choices, while not designed for the specific decision context, often provide reasonable and interpretable summaries of the posterior distribution, especially when the posterior is symmetric and unimodal. Moreover, they align closely with classical statistical estimators, facilitating comparisons between Bayesian and frequentist methods.

\subsection{Squared error loss}

A widely used and conceptually appealing loss function is the \emph{squared error loss}, defined as
\[
L(\boldsymbol{a},\boldsymbol{\theta}) = \| \boldsymbol{a} - \boldsymbol{\theta} \|_2^2 = \sum_{k=1}^K (a_k - \theta_k)^2,
\]
where \( \boldsymbol{a} \) is the action taken to estimate the unknown parameter \( \boldsymbol{\theta} \). This loss is symmetric around \( \boldsymbol{\theta} \) and penalizes errors quadratically, assigning disproportionately high loss to estimates far from the true value. In the frequentist literature, squared error loss leads to the familiar mean squared error criterion. From a Bayesian perspective, the optimal estimator under squared error loss is the posterior mean. That is,
\[
\hat{\boldsymbol{\theta}} = \arg \min_{\boldsymbol{a}} \int_\Theta \| \boldsymbol{a} - \boldsymbol{\theta} \|_2^2 \, p(\boldsymbol{\theta} \mid \boldsymbol{y}) \, \textsf{d}\boldsymbol{\theta} = \textsf{E}(\boldsymbol{\theta} \mid \boldsymbol{y}),
\]
where $\textsf{E}(\boldsymbol{\theta} \mid \boldsymbol{y})$ is the expectation of $\boldsymbol{\theta}$ under the posterior distribution $p(\boldsymbol{\theta}\mid\boldsymbol{y})$.

In the univariate case, this result follows from a simple decomposition of the squared error. For any \( a \in \mathbb{R} \),
\[
\int_\Theta (a - \theta)^2 \, p(\theta \mid \boldsymbol{y}) \, \textsf{d}\theta = \int_\Theta \left( a - \textsf{E}(\theta \mid \boldsymbol{y}) + \textsf{E}(\theta \mid \boldsymbol{y}) - \theta \right)^2 \, p(\theta \mid \boldsymbol{y}) \, \textsf{d}\theta.
\]
Expanding the square and using linearity of expectation, we obtain
\[
\int_\Theta (a - \theta)^2 \, p(\theta \mid \boldsymbol{y}) \, \textsf{d}\theta = (a - \textsf{E}(\theta \mid \boldsymbol{y}))^2 + \textsf{Var}(\theta \mid \boldsymbol{y}),
\]
where the second term does not depend on \( a \), and the first is minimized when \( a = \textsf{E}(\theta \mid \boldsymbol{y}) \).

Despite its optimality under squared loss, the posterior mean has several limitations. First, under proper priors, the posterior mean is generally a biased estimator of \( \theta \). Although to many statisticians unbiasedness is not a transcendental property, in fact, Bayesian estimators are typically biased \cite{casella2024statistical}. Second, the posterior mean is not invariant under transformations. That is, in general,
$\textsf{E}(h(\theta) \mid \boldsymbol{y}) \neq h(\textsf{E}(\theta \mid \boldsymbol{y}))$, which implies that if one is interested in a nonlinear functional \( h(\theta) \), it is necessary to compute the posterior expectation of \( h(\theta) \) directly, rather than applying the transformation to the posterior mean of \( \theta \).

Finally, a desirable property of the posterior mean is its \emph{invariance to marginalization}: it yields the same result whether computed by integrating over the joint posterior or over the marginal distribution of interest. While exact expressions for posterior means are sometimes tractable when the posterior belongs to a conjugate family, they can become difficult to obtain in closed form for complex models or nonlinear functionals. In such cases, simulation-based methods such as Markov Chain Monte Carlo (MCMC; e.g., \citealt{gamerman2006markov}) are often used to approximate posterior expectations.

\subsection{Absolute error loss}

An important alternative to the squared error loss is the \emph{absolute error loss}, defined as
\[
L(\boldsymbol{a},\boldsymbol{\theta}) = \| \boldsymbol{a} - \boldsymbol{\theta} \|_1 = \sum_{k=1}^K |a_k - \theta_k|,
\]
where \( \boldsymbol{a} \) is the action taken to estimate the unknown parameter \( \boldsymbol{\theta} \). Like squared error loss, this loss function is symmetric around the true value, but it penalizes deviations linearly rather than quadratically, making it less sensitive to outliers. The Bayes estimator under absolute error loss is given by
\[
\hat{\boldsymbol{\theta}} = \arg \min_{\boldsymbol{a}} \int_\Theta \| \boldsymbol{a} - \boldsymbol{\theta} \|_1 \, p(\boldsymbol{\theta} \mid \boldsymbol{y}) \, \textsf{d}\boldsymbol{\theta}.
\]
The solution to this problem is the posterior median of each component \( \theta_k \), defined implicitly by
\[
\int_{-\infty}^{\hat{\theta}_k} p(\theta_k \mid \boldsymbol{y}) \, \textsf{d}\theta_k = \frac{1}{2}, \qquad \text{for } k = 1, \ldots, K.
\]

In contrast to the posterior mean, the posterior median shares two desirable properties with the maximum likelihood estimator: it is invariant under monotone transformations, and it is invariant to marginalization. Moreover, the absolute error loss framework can be easily generalized to allow for asymmetric penalties. This flexibility allows practitioners to adapt the loss function to emphasize overestimation or underestimation, depending on the specific application.

\subsection{0-1 error loss}

A widely used utility function, particularly relevant for discrete parameter spaces—is the \emph{0–1 utility}, defined as
\[
U(\theta, a) =
\begin{cases}
1, & \text{if } a = \theta, \\
0, & \text{otherwise},
\end{cases}
\]
where \( a \) is the action taken to estimate \( \theta \). Unlike loss functions, which quantify the penalty associated with incorrect decisions, utility functions represent rewards and are maximized rather than minimized. In decision-theoretic terms, using a utility function is especially natural when the goal is to model success rather than cost. The 0–1 utility assigns a reward of 1 only when the estimate exactly equals the true value and 0 otherwise. The Bayes rule under this utility is the action that maximizes the posterior expected utility, which in this case corresponds to selecting the value of \( \theta \) with the highest posterior probability, i.e., the \emph{posterior mode}:
\[
\hat{\theta} = \arg\max_{\theta} p(\theta \mid \boldsymbol{y}).
\]

The idea can be extended to continuous parameter spaces using a limiting argument. Let \( \text{B}_{\theta, \Delta} \) denote a ball of diameter \( \Delta > 0 \) centered at \( \theta \), and define the approximate utility function
\[
U_\Delta(a, \theta) =
\begin{cases}
1, & \text{if } a \in \text{B}_{\theta, \Delta}, \\
0, & \text{otherwise}.
\end{cases}
\]
The Bayes estimator under this utility is given by
\[
\hat{\theta}_\Delta = \arg\max_a \textsf{E}(U_\Delta(a, \theta) \mid \boldsymbol{y}) = \arg\max_a \int_\Theta U_\Delta(a, \theta) \, p(\theta \mid \boldsymbol{y}) \, \textsf{d}\theta,
\]
and taking the limit as \( \Delta \to 0 \) yields the posterior mode $\hat{\theta} = \lim_{\Delta \to 0} \hat{\theta}_\Delta$.

The posterior median is invariant under monotone transformations (in 1D). In contrast, MAP/posterior modes generally depend on parameterization. However, unlike the posterior mean, it is not invariant under marginalization: computing the mode of a marginal posterior distribution may yield a different result than marginalizing the full posterior mode. This behavior contrasts with the posterior mean, which is invariant to marginalization but not to nonlinear transformations.

\section{Bayesian interval estimation}

In Bayesian statistics, interval estimation is carried out by means of \emph{credible sets}. Given a parameter \( \theta \), a \( 100(1 - \alpha)\% \) credible set \( \text{C}_\alpha \subset \Theta \) is defined as a subset of the parameter space \( \Theta \) such that
\[
\int_{\text{C}_\alpha} p(\theta \mid \boldsymbol{y}) \, \textsf{d}\theta = 1 - \alpha.
\]
In other words, the posterior probability that \( \theta \) lies within \( \text{C}_\alpha \) is equal to \( 1 - \alpha \).

Unlike the frequentist confidence interval, which is defined by its long-run coverage properties under repeated sampling, a Bayesian credible set provides a direct probabilistic statement about the parameter conditional on the observed data. However, for any fixed level \( \alpha \), the credible set \( \text{C}_\alpha \) is not unique: there are infinitely many subsets of the parameter space that satisfy the posterior probability condition. This raises an important question: how should one choose among the many possible credible sets? The answer depends on the inferential goals and, in a fully Bayesian framework, should be guided by a utility function defined over subsets of the parameter space. The appropriate utility function will depend on the context of the problem and the consequences associated with including or excluding particular values of \( \theta \). Choosing a credible set can therefore be framed as a decision problem, where the optimal set maximizes expected utility under the posterior distribution.

\subsection{Highest posterior density intervals}

A desirable property of any credible set is that it should be as small as possible while still containing the parameter with high posterior probability. One principled way to formalize this trade-off is through a loss function defined over credible sets \( \text{C}_\alpha \subset \Theta \), given by
\[
L_k(\text{C}_\alpha, \theta) = k \, \mathbb{I}(\theta \notin \text{C}_\alpha) + |\text{C}_\alpha|,
\]
where \( \mathbb{I}_A(\cdot) \) denotes the indicator function of the set \( A \), \( |A| \) represents the length or volume of the set \( A \), and \( k > 0 \) controls the relative weight assigned to the cost of excluding the true parameter value. The first term penalizes credible sets that fail to contain \( \theta \), while the second penalizes excessive width or volume, thereby promoting more precise and informative intervals.

The problem then becomes one of minimizing the expected loss subject to a posterior probability constraint:
\[
\int_{\text{C}_\alpha} p(\theta \mid \boldsymbol{y}) \, \textsf{d}\theta = 1 - \alpha.
\]
The optimal solution under this framework is the \emph{highest posterior density (HPD) set}, which is defined as the smallest subset of the parameter space that contains posterior probability \( 1 - \alpha \). Formally, the HPD set includes all values of \( \theta \) for which the posterior density exceeds a threshold \( c_\alpha \) chosen to satisfy the probability constraint:
\[
\text{C}_\alpha^{\text{HPD}} = \left\{ \theta \in \Theta : p(\theta \mid \boldsymbol{y}) \geq c_\alpha \right\}, \quad \text{where} \quad \int_{\text{C}_\alpha^{\text{HPD}}} p(\theta \mid \boldsymbol{y}) \, \textsf{d}\theta = 1 - \alpha.
\]

An important consequence of the HPD definition is that the posterior density is constant along the boundary of the set. This property suggests a simple graphical procedure for constructing HPD sets in the univariate case: plot the posterior density and gradually lower a horizontal line from \( +\infty \) until the total area under the regions where the density exceeds the line equals \( 1 - \alpha \). The resulting region where the posterior density is above this threshold defines the HPD set. It is important to note that HPD sets are not necessarily contiguous. When the posterior distribution is multimodal, the HPD region may consist of multiple disjoint intervals, each surrounding a local mode. This behavior contrasts with central credible intervals, which are always connected but may include areas of low posterior density between modes.

HPD intervals may also include the boundary of the parameter space, particularly when the posterior is concentrated near the extremes. For instance, consider the model \( y \mid \theta \sim \textsf{Bin}(n, \theta) \), with a uniform prior \( \theta \sim \textsf{Uniform}[0,1] \). If the observed data are \( y = 0 \) or \( y = n \), the posterior becomes highly skewed and concentrated near the boundaries \( \theta = 0 \) or \( \theta = 1 \), respectively. In such cases, the HPD interval will include the endpoint of the support. Furthermore, while HPD intervals are relatively straightforward to compute for univariate, symmetric posterior distributions, they can become computationally challenging in more complex settings. This is particularly true when the posterior is asymmetric, multimodal, or defined over constrained spaces, or when its normalization constant and cumulative integrals cannot be evaluated in closed form. In such cases, numerical integration or simulation-based methods (e.g., MCMC) are required to approximate HPD regions.

\subsection{Equal-tailed credible intervals}

In the continuous univariate setting, a common approach to constructing credible intervals is based on specifying an explicit utility function over intervals. Let \( \textsf{C}_\alpha = (a_\alpha, b_\alpha) \subset \mathbb{R} \) denote a credible interval of level \( 1 - \alpha \). One possible utility function that balances the inclusion of the true parameter value with the symmetry and simplicity of the interval is given by
\begin{align*}
L_\alpha(\theta, (a_\alpha, b_\alpha)) &= \tfrac{\alpha}{2} (a_\alpha - \theta) \, \mathbb{I}(\theta < a_\alpha) 
+ \left(1 - \tfrac{\alpha}{2} \right) (\theta - a_\alpha) \, \mathbb{I}(\theta \geq a_\alpha) \\
&\hspace{3.8cm}\, + \left(1 - \tfrac{\alpha}{2} \right) (b_\alpha - \theta) \, \mathbb{I}(\theta < b_\alpha) 
+ \tfrac{\alpha}{2} (\theta - b_\alpha) \, \mathbb{I}(\theta \geq b_\alpha),
\end{align*}
subject to the constraint
\[
\int_{a_\alpha}^{b_\alpha} p(\theta \mid \boldsymbol{y}) \, \textsf{d}\theta = 1 - \alpha.
\]

This utility function favors credible intervals that allocate equal posterior probability to both tails of the distribution, resulting in what is known as the \emph{equal-tailed credible interval}. This interval \( (a_\alpha, b_\alpha) \) is defined by the condition
\[
\int_{-\infty}^{a_\alpha} p(\theta \mid \boldsymbol{y}) \, \textsf{d}\theta = \int_{b_\alpha}^{\infty} p(\theta \mid \boldsymbol{y}) \, \textsf{d}\theta = \frac{\alpha}{2}.
\]
Equal-tailed intervals are typically easier to compute than HPD intervals, particularly when the posterior distribution is available in closed form or when quantile-based approximations can be employed. In the case of symmetric and unimodal posterior distributions, the equal-tailed interval coincides with the HPD interval. However, for skewed or multimodal posteriors, the two may differ substantially. The choice between them should take into account both computational convenience and the interpretive priorities of the analysis.

\subsection*{Frequentist coverage}

A foundational result by \citet{hartigan1966note} establishes that standard Bayesian credible intervals possess desirable frequentist properties in large samples. Specifically, let \( [L(\boldsymbol{y}), U(\boldsymbol{y})] \) denote a \( 100(1 - \alpha)\% \) Bayesian credible interval for a scalar parameter \( \theta \). Hartigan showed that this interval satisfies the following frequentist coverage property:
\[
\Pr\left( L(\boldsymbol{y}) < \theta < U(\boldsymbol{y}) \mid \theta \right) = (1 - \alpha) + \epsilon_n,
\]
where the probability is taken over repeated samples \( \boldsymbol{y} \sim p(\boldsymbol{y} \mid \theta) \), \( \epsilon_n \to 0 \) as \( n \to \infty \), and \( |\epsilon_n| < \frac{a}{n} \) for some constant \( a > 0 \). In other words, Bayesian credible intervals achieve asymptotically correct frequentist coverage, matching the behavior of confidence intervals derived from large-sample frequentist methods.

This result reinforces the frequentist validity of Bayesian procedures in large samples and provides a theoretical justification for using credible intervals beyond purely subjective interpretation. In fact, some authors have argued that credible intervals derived from fully Bayesian reasoning may offer a more principled and reliable alternative to intervals obtained from asymptotic approximations, particularly when the latter rely on conditions that may not hold in practice.

\subsection{Example: Binomial-Beta model}

Consider a model where \( y \mid \theta \sim \textsf{Bin}(n, \theta) \), and assume a uniform prior \( \theta \sim \textsf{Uniform}[0, 1] \). If the observed data are \( y = 0 \), then the posterior distribution of \( \theta \) is given by \( \theta \mid y \sim \textsf{Beta}(1, n + 1) \). To derive the HPD interval, observe that the posterior density is \( p(\theta \mid y) = (n+1)(1 - \theta)^n \) for \( \theta \in [0, 1] \), which is strictly decreasing. Consequently, the HPD interval takes the form \( (0, b) \), where \( b \) is chosen to satisfy the posterior probability constraint:
\[
\int_0^b (n+1)(1 - \theta)^n \, \textsf{d}\theta = 1 - \alpha.
\]
Using the substitution \( u = 1 - \theta \), this becomes
\[
\int_0^b (n+1)(1 - \theta)^n \, \textsf{d}\theta 
= \int_{1 - b}^1 (n+1) u^n \, \textsf{d}u 
= 1 - (1 - b)^{n+1}.
\]
Equating to \( 1 - \alpha \), we find \( (1 - b)^{n+1} = \alpha \), which implies $b = 1 - \alpha^{1/(n+1)}$. Thus, the HPD interval is given by \( \big( 0, \, 1 - \alpha^{1/(n+1)} \big) \).

To derive the equal-tailed credible interval for \( \theta \mid y = 0 \sim \textsf{Beta}(1, n+1) \), note that its cumulative distribution function is \( F(\theta) = 1 - (1 - \theta)^{n+1} \). The equal-tailed \( 100(1 - \alpha)\% \) credible interval \( (a, b) \) satisfies:
$F(a) = \frac{\alpha}{2}$ and $F(b) = 1 - \frac{\alpha}{2}$. Solving for \( a \) and \( b \), we get that the equal-tailed credible interval is given by $\big(1 - \left(1 - \alpha/2 \right)^{1/(n+1)}, \, 1 - \left( \alpha/2 \right)^{1/(n+1)} \big)$.

For example, if \( n = 10 \) and \( \alpha = 0.95 \), the HPD interval is \( (0,\, 0.20553) \), while the equal-tailed interval is \( (0.00775,\, 0.26724) \). This illustrates a case where the standard frequentist confidence interval is undefined due to the boundary observation \( y = 0 \). In contrast, the Bayesian approach provides valid and interpretable intervals, even when the data lie at the edge of the parameter space.

\section{Bayesian hypothesis testing}

Bayesian hypothesis testing can be naturally framed as a decision problem over a discrete parameter space. Specifically, the task of comparing models can be viewed as estimating a discrete latent variable that indexes competing hypotheses. Consider a hierarchical framework involving \( K \) candidate models \( \textsf{M}_1, \ldots, \textsf{M}_K \), where each model \( \textsf{M}_k \) specifies a likelihood \( p(\boldsymbol{y} \mid \theta_k, \textsf{M}_k) \), a prior distribution over the model-specific parameters \( p(\theta_k \mid \textsf{M}_k) \), and a prior probability for the model itself \( p(\textsf{M}_k) \). The primary quantity of interest is the posterior probability of each model:
\[
p(\textsf{M}_k \mid \boldsymbol{y}) \propto p(\boldsymbol{y} \mid \textsf{M}_k) \, p(\textsf{M}_k) = p(\textsf{M}_k) \int_\Theta p(\boldsymbol{y}, \theta_k \mid \textsf{M}_k) \, \textsf{d}\theta_k = p(\textsf{M}_k) \int_\Theta p(\boldsymbol{y} \mid \theta_k, \textsf{M}_k) \, p(\theta_k \mid \textsf{M}_k) \, \textsf{d}\theta_k,
\]
which integrates out the model-specific parameters \( \theta_k \) and therefore involves the marginal likelihood (prior predictive distribution) under each model \( \textsf{M}_k \).

Unlike frequentist hypothesis testing, which is typically restricted to pairwise comparisons and often relies on asymptotic approximations, the Bayesian approach allows for the simultaneous comparison of multiple competing hypotheses. Since the model space \( \{\textsf{M}_1, \ldots, \textsf{M}_K\} \) is discrete, it is natural to adopt a 0–1 utility function, analogous to that used in point estimation for discrete parameters, which assigns utility 1 for selecting the true model and 0 otherwise. Under this utility, the Bayes-optimal decision rule is to select the model with the highest posterior probability:
\[
\tilde{k} = \arg\max_k \, p(\textsf{M}_k \mid \boldsymbol{y}).
\]
This decision-theoretic framework offers a coherent and interpretable alternative to classical hypothesis testing, while also providing a principled way to account for uncertainty in model selection through the posterior distribution over models.

\subsection*{Bayes factors and posterior odds}

The posterior probability of model \( \textsf{M}_k \) can be conveniently expressed in terms of model odds. Specifically, by multiplying and dividing by \( p(\boldsymbol{y} \mid \textsf{M}_k)\, p(\textsf{M}_k) \), we obtain
\[
p(\textsf{M}_k \mid \boldsymbol{y}) = \frac{p(\boldsymbol{y} \mid \textsf{M}_k)\, p(\textsf{M}_k)}{\sum_{\ell=1}^K p(\boldsymbol{y} \mid \textsf{M}_\ell)\, p(\textsf{M}_\ell)} = \frac{1}{1 + \sum_{\ell:\ell \neq k} o_{\ell,k}},
\]
where \( o_{\ell,k} \) denotes the posterior odds of model \( \textsf{M}_\ell \) against model \( \textsf{M}_k \), given by
\[
o_{\ell,k} = \frac{p(\textsf{M}_\ell \mid \boldsymbol{y})}{p(\textsf{M}_k \mid \boldsymbol{y})} = \frac{p(\textsf{M}_\ell)}{p(\textsf{M}_k)} \cdot \frac{p(\boldsymbol{y} \mid \textsf{M}_\ell)}{p(\boldsymbol{y} \mid \textsf{M}_k)}.
\]
Here, the first term \( p(\textsf{M}_\ell)/p(\textsf{M}_k) \) is the \emph{prior odds}, and the second term
\[
\text{BF}_{\ell,k} = \frac{p(\boldsymbol{y} \mid \textsf{M}_\ell)}{p(\boldsymbol{y} \mid \textsf{M}_k)}
\]
is the \emph{Bayes factor}, which quantifies the relative support the data provide for model \( \textsf{M}_\ell \) over model \( \textsf{M}_k \). Thus, Bayes factors are the change in the prior odds after observing the data. This clarifies the correct interpretation of Bayes factors because there have been some misconceptions. An excellent discussion on the matter can be found in \citet{lavine1999bayes}. 

While some practitioners prefer to express model comparisons using log-odds or log Bayes factors (for reasons of interpretability or numerical stability) the information they convey is equivalent to working directly with posterior model probabilities. This formulation grounds hypothesis testing and model selection firmly in probability theory, avoiding reliance on asymptotic approximations, $p$-values, or arbitrary significance thresholds. However, the interpretation of posterior probabilities and Bayes factors is not always straightforward. \citet{kass1995bayes} proposed a widely adopted set of guidelines, as summarized in Table~\ref{tab:bayes_factors}, for assessing the strength of evidence in favor of one model over another, based on posterior odds or Bayes factors. Lastly, it is important to emphasize that, unlike estimation problems, Bayesian and frequentist approaches to hypothesis testing can lead to substantially different conclusions, especially in finite samples or under model misspecification.

\begin{table}[!htb]
\centering
\begin{tabular}{lll}
\toprule
Posterior Odds (against) & Posterior Probability & Interpretation \\
\midrule
1 to 3    & 0.50 to 0.75 & Not worth more than a bare mention \\
3 to 20   & 0.75 to 0.95 & Positive evidence \\
20 to 150 & 0.95 to 0.99 & Strong evidence \\
$>$150    & $>$0.99      & Very strong evidence \\
\bottomrule
\end{tabular}
\caption{Guidelines for interpreting posterior odds and Bayes factors, as proposed by \citet{kass1995bayes}.}
\label{tab:bayes_factors}
\end{table}

\subsection{Example: Lindley’s paradox}

Consider a city in which 49,581 boys and 48,870 girls were born over a certain time period. Assuming each birth is independent and that the probability of a male birth is constant, the total number of boys \( y \) can be modeled as a binomial random variable $y \mid \theta \sim \textsf{Bin}(n = 98{,}451,\, \theta)$, where \( \theta \) is the probability of a male birth. We are interested in testing the null hypothesis \( H_0 : \theta = 0.5 \) versus the two-sided alternative \( H_1 : \theta \neq 0.5 \).

Using a frequentist approach, we test \( H_0 : \theta = 0.5 \) by applying a normal approximation to the binomial sampling distribution. Under \( H_0 \), the number of male births \( y \sim \textsf{Bin}(n = 98{,}451,\, \theta = 0.5) \) has expected value \( \textsf{E}(y) = n\theta = 49{,}225.50 \) and variance \( \textsf{Var}(y) = n\theta(1 - \theta) = 24{,}612.75 \). Approximating the distribution of \( y \) with a normal distribution, we compute the standardized test statistic, which quantifies how many standard deviations the observed count deviates from the null mean. The corresponding two-tailed $p$-value is
\[
\text{$p$-value} = \Pr\left( \frac{Y - n\theta}{\sqrt{n\theta(1 - \theta)}} \geq \frac{49{,}581 - 49{,}225.50}{\sqrt{24{,}612.75}} \,\middle|\, \theta = 0.5 \right) = \Pr(Z \geq 2.26600) \approx 0.02345,
\]
where \( Z \) is a standard normal variable. Since this value falls below the conventional 5\% threshold, the classical test rejects \( H_0 \), indicating a statistically significant deviation from the hypothesized proportion \( \theta = 0.5 \).

In the Bayesian framework, model comparison relies on the marginal likelihood, which averages the likelihood function over the prior distribution specific to each hypothesis. To compare the models \( H_0 \colon \theta = 0.5 \) and \( H_1 \colon \theta \in (0,1) \), with \( \theta \sim \textsf{Uniform}(0,1) \) under \( H_1 \), we compute the marginal likelihood under each. Under \( H_0 \), the parameter \( \theta \) is fixed at 0.5, so the marginal likelihood is the probability of observing the data under a binomial model with that fixed value:
\[
p(y \mid H_0) = \binom{n}{y} \cdot 0.5^y \cdot (1 - 0.5)^{n - y} = \binom{n}{y} \cdot 0.5^n \approx 1.95 \times 10^{-4}.
\]
Under \( H_1 \), the parameter \( \theta \) is unknown and assumed to follow a uniform prior over the interval \([0,1]\). The marginal likelihood is obtained by integrating the likelihood over this prior:
\[
p(y \mid H_1) = \int_0^1 p(y \mid \theta) \,p(\theta \mid H_1) \, \textsf{d}\theta 
= \binom{n}{y} \int_0^1 \theta^y (1 - \theta)^{n - y} \, \textsf{d}\theta 
= \binom{n}{y} B(y + 1, n - y + 1) \approx 1.02 \times 10^{-5},
\]
where \( B(\cdot, \cdot) \) denotes the Beta function. Assuming equal prior probabilities for the two hypotheses, \( p(H_0) = p(H_1) = 0.5 \), the posterior probability of the null hypothesis is given by
\[
p(H_0 \mid y) = \frac{p(y \mid H_0) \cdot p(H_0)}{p(y \mid H_0) \cdot p(H_0) + p(y \mid H_1) \cdot p(H_1)} 
= \frac{p(y \mid H_0)}{p(y \mid H_0) + p(y \mid H_1)} \approx 0.95.
\]
That is, despite the frequentist test rejecting \( H_0 \) at the 5\% significance level, the Bayesian analysis assigns high posterior probability to the null, thereby favoring it. This divergence in conclusions illustrates a classic example of \emph{Lindley’s paradox}.

To further understand this discrepancy, consider the power of the frequentist test, i.e., the probability of correctly rejecting the null hypothesis when it is false. In particular, evaluate the power to detect a deviation as small as the one actually observed, which corresponds to \( \theta = 49{,}581 / 98{,}451 \approx 0.50361 \). Using the standard normal approximation, the power is given by
\[
1 - \Pr\left( -1.95996 < \frac{y - n\theta}{\sqrt{n\theta(1 - \theta)}} < 1.95996 \,\middle|\, \theta = 0.50361 \right) \approx 0.62002.
\]
This indicates that, even if the true proportion of male births were 0.50360, the classical test would correctly reject the null only about 62\%, reflecting the limited sensitivity of the test to such a small deviation from the hypothesized value. 

In the Bayesian framework, hypothesis testing is often conducted using the Bayes factor, defined as \( B_{01} = p(y \mid H_0) / p(y \mid H_1) \), where \( H_0 \) fixes \( \theta = 0.5 \), and \( H_1 \) assigns a uniform prior over \( \theta \in (0,1) \). The Bayes factor quantifies the relative evidence the data provide for the null hypothesis compared to the alternative. A common decision rule is to favor \( H_1 \) when \( B_{01} < 1 \), indicating that the data are more likely under the alternative model. Within this framework, the Bayesian analogue of the Type I error is \( \Pr(B_{01} < 1 \mid \theta = 0.5) \approx 0.00093 \), which can be estimated via Monte Carlo simulation by repeatedly sampling data from the binomial distribution under \( \theta = 0.5 \), computing the Bayes factor for each replicate, and evaluating the proportion for which \( B_{01} < 1 \). This extremely low false positive rate reflects the inherent conservatism of the Bayesian test, which demands strong evidence to reject the null. In contrast, the power of the Bayesian test is \( \Pr(B_{01} < 1 \mid \theta = 0.50361) \approx 0.14516 \), substantially lower than that of the corresponding frequentist test. This illustrates a fundamental trade-off in Bayesian hypothesis testing: its resilience against false positives comes at the expense of reduced sensitivity to small deviations from the null. This behavior stems from the integration over the entire parameter space under \( H_1 \), in contrast to the frequentist approach, which controls Type I error explicitly but may react more strongly to marginal departures from the null.

These results demonstrate that the Bayesian test tends to be considerably more conservative in the presence of small deviations from the null, especially when prior mass is concentrated under \( H_0 \). This behavior encapsulates the core insight of Lindley’s paradox: even when data appear statistically significant under frequentist criteria, the Bayesian analysis may continue to favor the null due to the influence of prior structure and marginal likelihood. $\hfill\square$

\subsection{Example: Normal-Normal model}

When both hypotheses in a testing problem are composite, Bayes factors can often be naturally re-expressed in terms of posterior probabilities. Consider the model \( y_i \mid \theta \overset{\text{i.i.d.}}{\sim} \textsf{N}(\theta, 1) \) for \( i = 1, \ldots, n \), and suppose we wish to test the one-sided hypotheses \( H_0 \colon \theta \leq \theta_0 \) versus \( H_1 \colon \theta > \theta_0 \). To construct priors under each hypothesis, we begin with a common Gaussian prior \( \theta \sim \textsf{N}(\mu, \tau^2) \) and truncate it appropriately. The resulting conditional priors are:
\[
p(\theta \mid H_0) = \frac{1}{\sqrt{2\pi\tau^2} \, \Phi\left(\frac{\theta_0 - \mu}{\tau}\right)} \, \exp\left\{ -\frac{1}{2\tau^2} (\theta - \mu)^2 \right\} \cdot \mathbb{I}(\theta \leq \theta_0),
\]
and
\[
p(\theta \mid H_1) = \frac{1}{\sqrt{2\pi\tau^2} \, \left[1 - \Phi\left(\frac{\theta_0 - \mu}{\tau}\right)\right]} \, \exp\left\{ -\frac{1}{2\tau^2} (\theta - \mu)^2 \right\} \cdot \mathbb{I}(\theta > \theta_0),
\]
where \( \Phi(\cdot) \) denotes the cumulative distribution function of the standard normal distribution. These expressions correspond to truncated normal priors under each hypothesis, ensuring that the prior support aligns with the hypothesis being tested.

Under these priors, the posterior odds in favor of \( H_0 \) over \( H_1 \) take the form:
\[
O_{01} = 
\frac{\epsilon}{1 - \epsilon} \cdot 
\frac{\int_{-\infty}^{\theta_0} \exp\left\{-\frac{1}{2} \left(\frac{1}{\tau^2} + n\right) \left(\theta - \frac{\mu/\tau^2 + n\bar{y}}{1/\tau^2 + n} \right)^2 \right\} \, \textsf{d}\theta}
{\int_{\theta_0}^{\infty} \exp\left\{-\frac{1}{2} \left(\frac{1}{\tau^2} + n\right) \left(\theta - \frac{\mu/\tau^2 + n\bar{y}}{1/\tau^2 + n} \right)^2 \right\} \, \textsf{d}\theta},
\]
where \( \bar{y} = \frac{1}{n} \sum_{i=1}^n y_i \) is the sample mean, and \( \epsilon = p(H_0) = 1 - p(H_1) \) is the prior probability assigned to the null hypothesis. This expression makes clear that the posterior odds are the product of the prior odds and the ratio of posterior mass assigned to the parameter regions consistent with each hypothesis.

Importantly, if we choose \( \epsilon = \Phi\left(\frac{\theta_0 - \mu}{\tau}\right) \), that is, we assign prior model probabilities proportional to the prior mass allocated to each hypothesis, then the posterior odds reduce to
\[
O_{0,1} = \frac{\Pr(\theta \leq \theta_0 \mid \boldsymbol{y})}{\Pr(\theta > \theta_0 \mid \boldsymbol{y})},
\]
which is simply the ratio of posterior probabilities in the regions consistent with \( H_0 \) and \( H_1 \), respectively. Similarly, if we instead choose \( \mu = \theta_0 \), thereby centering the prior symmetrically at the null boundary, then \( \Phi\left(\frac{\theta_0 - \mu}{\tau}\right) = 0.5 \), and the posterior odds take the form
\[
O_{0,1} = \frac{\epsilon}{1 - \epsilon} \cdot \frac{\Pr(\theta \leq \theta_0 \mid \boldsymbol{y})}{\Pr(\theta > \theta_0 \mid \boldsymbol{y})}.
\]

These expressions show that, with symmetric and well-calibrated priors, Bayes factors for composite hypotheses can be interpreted directly as ratios of posterior probabilities, bypassing the need for explicit marginal likelihood computations. More broadly, the example illustrates the complexity of prior elicitation in Bayesian testing: the effective prior assigned to each model depends not only on the model probabilities \( \epsilon \) and \( 1 - \epsilon \), but also on the location \( \mu \) and scale \( \tau^2 \) of the underlying prior for \( \theta \). Even when using a common base distribution, the truncation and renormalization required by each hypothesis can yield substantially different priors, influencing the resulting inference. Thus, careful prior specification is crucial to ensure that posterior comparisons faithfully reflect both prior beliefs and data evidence. $\hfill\square$

\section{Prediction}

In the Bayesian framework, both point and interval prediction can be naturally treated as estimation problems. Optimal prediction rules are obtained by minimizing expected loss with respect to the posterior predictive distribution, using utility functions analogous to those employed in parameter estimation.

For point prediction, a common choice is to minimize the squared error loss. The optimal prediction \( \tilde{y}^* \) is then defined as
\[
\tilde{y}^* = \arg\min_{a} \int_{\mathcal{Y}} (a - y^*)^2 \, p(y^* \mid \boldsymbol{y}) \, \textsf{d}y^*,
\]
where \( \mathcal{Y} \) denotes the sample space of possible future observations, and \( p(y^* \mid \boldsymbol{y}) \) is the posterior predictive distribution. The solution to this problem is the posterior predictive mean
\[
\tilde{y}^* = \textsf{E}(y^* \mid \boldsymbol{y}) = \int_{\mathcal{Y}} y^* \, p(y^* \mid \boldsymbol{y}) \, \textsf{d}y^*.
\]

When multiple models \( \{ \textsf{M}_1, \ldots, \textsf{M}_K \} \) are under consideration, prediction can be extended using Bayesian model averaging. In this setting, the optimal point prediction integrates over both parameter and model uncertainty. Under squared error loss, the Bayes estimator is the posterior predictive mean, \( \tilde{y}^* = \textsf{E}(y^* \mid \boldsymbol{y}) \). Applying the law of total probability, the posterior predictive distribution can be decomposed as
\[
p(y^* \mid \boldsymbol{y}) = \sum_{k=1}^K p(y^*, \textsf{M}_k \mid \boldsymbol{y}) = \sum_{k=1}^K p(y^* \mid \boldsymbol{y}, \textsf{M}_k) \cdot p(\textsf{M}_k \mid \boldsymbol{y}),
\]
which implies
\[
\textsf{E}(y^* \mid \boldsymbol{y}) = \int_{\mathcal{Y}} y^* \, p(y^* \mid \boldsymbol{y}) \, \textsf{d}y^* = \int_{\mathcal{Y}} y^* \left( \sum_{k=1}^K p(y^* \mid \boldsymbol{y}, \textsf{M}_k) \cdot p(\textsf{M}_k \mid \boldsymbol{y}) \right) \, \textsf{d}y^*.
\]
By linearity of expectation and Fubini’s theorem (interchanging sum and integral), we obtain
\[
\textsf{E}(y^* \mid \boldsymbol{y}) = \sum_{k=1}^K p(\textsf{M}_k \mid \boldsymbol{y}) \cdot \int_{\mathcal{Y}} y^* \, p(y^* \mid \boldsymbol{y}, \textsf{M}_k) \, \textsf{d}y^* = \sum_{k=1}^K \textsf{E}(y^* \mid \boldsymbol{y}, \textsf{M}_k) \cdot p(\textsf{M}_k \mid \boldsymbol{y}).
\]
This expression shows that the optimal Bayesian prediction is a weighted average of model-specific predictions, where the weights are given by the posterior model probabilities. It provides a coherent and principled approach to account for model uncertainty in predictive inference.

A similar strategy applies to interval prediction. Credible intervals for future observations can be derived directly from the posterior predictive distribution, yielding probabilistic statements about \( y^* \) that incorporate uncertainty in both the model parameters and the model selection itself.

\section{Bayesian identifiability}

Identifiability plays an important role in statistical inference. Under the frequentist paradigm, it does not make sense to infer nonidentifiable parameters. However, many data analysts overlook this relevant concept and carry out hypothesis tests or report interval estimates of non-identifiable parameters. \citet{dawid1979conditional} gave the following formal definition of identifiability in the Bayesian framework. Consider a Bayesian model with likelihood $p(\boldsymbol{y}\mid \boldsymbol{\theta})$ and prior $p(\boldsymbol{\theta})$, and a partition of the the parameter as $\boldsymbol{\theta} = (\boldsymbol{\theta}_1,\boldsymbol{\theta}_2)$. Then $\boldsymbol{\theta}_2$ is said to be nonidentifiable if it is conditionally independent of $\boldsymbol{y}$ given $\boldsymbol{\theta}_1$, that is,
\begin{align*}
\boldsymbol{\theta}_2 \perp \boldsymbol{y} \mid \boldsymbol{\theta}_1
\quad \Longleftrightarrow \quad
p(\boldsymbol{\theta}_2 \mid \boldsymbol{\theta}_1,\boldsymbol{y})
=
p(\boldsymbol{\theta}_2 \mid \boldsymbol{\theta}_1).
\end{align*}
Therefore, given $\boldsymbol{\theta}_1$, the data do not increase prior knowledge about $\boldsymbol{\theta}_2$. It turns out that a necessary and sufficient condition for the non-identifiability of $\boldsymbol{\theta}_2$ is that the likelihood does not depend on $\boldsymbol{\theta}_2$, namely,
\begin{align*}
p(\boldsymbol{y} \mid \boldsymbol{\theta}_1,\boldsymbol{\theta}_2)
=
p(\boldsymbol{y} \mid \boldsymbol{\theta}_1),
\end{align*}
as discussed by \citet{gelfand1999identifiability}. Moreover, \citet{poirier1998revising} showed that the frequentist definition of non-identifiability implies Bayesian non-identifiability. Differing from the frequentist scenario, in the Bayesian one, non-identifiability does not automatically imply that there is no Bayesian learning about $\boldsymbol{\theta}_2$, because it does not necessarily imply that
\begin{align*}
p(\boldsymbol{\theta}_2 \mid \boldsymbol{y}) = p(\boldsymbol{\theta}_2).
\end{align*}

\citet{lindley1971bayesian} discussed that posing objective priors over all model unknowns allows Bayesian analysis of non-identifiable models. However, in this case we can use the famous quote ``there is no Bayesian free lunch'', because the price that has to be paid is that for certain parameters there is no Bayesian learning, meaning that the data are uninformative for them \citep{poirier1998revising}. This also has consequences when using MCMC algorithms to perform inference \citep{gelfand1999identifiability}. Moreover, the problem of non-identifiability is more complicated when using improper priors \citep{poirier1998revising}.

When using flat improper priors, posterior propriety is an issue. In this regard, the following general result \citep{ghosh1998bayesian} has been used to find conditions for posterior propriety in particular models. Suppose that a Bayesian model has an unidentified likelihood such that
\begin{align*}
p(\boldsymbol{y} \mid \boldsymbol{\theta}_1,\boldsymbol{\theta}_2)
=
p(\boldsymbol{y} \mid \boldsymbol{\theta}_1),
\end{align*}
and an improper prior $p(\boldsymbol{\theta}_1,\boldsymbol{\theta}_2)$ is adopted. Then the joint posterior $p(\boldsymbol{\theta}_1,\boldsymbol{\theta}_2 \mid \boldsymbol{y})$ is proper if, and only if, the marginal posterior $p(\boldsymbol{\theta}_1 \mid \boldsymbol{y})$ and the conditional prior $p(\boldsymbol{\theta}_2 \mid \boldsymbol{\theta}_1)$ are proper.

For example, \citet{gelfand1999identifiability} used this lemma to derive conditions to check the propriety of the posterior for generalized linear models, with particular results for the special case of the usual Gaussian linear model. Furthermore, \citet{poirier1998revising} discussed Bayesian learning from a general perspective and provided several examples for particular cases such as hierarchical Gaussian linear models, censored sampling, and binary data. Currently, there are some approximations to check identifiability in complex models based on MCMC algorithms. In practice, lack of convergence, as well as wide or multimodal posteriors, suggest non-identifiability.

\section{Hierarchical modeling}

In the Bayesian framework, it is natural to treat the parameters of prior distributions as unknown and to assign them additional distributions, referred to as \textit{hyperpriors}. This leads to hierarchical, or multilevel, models in which uncertainty is structured across multiple layers (\citealt{lehmann1998theory},  \citealt{BernardoSmith2000}, and \citealt{GhoshEtAl2007}). This approach, also known as \textit{hierarchical Bayes} has the advantage of allowing the modeling of complex problems by decomposing its probabilistic representation into a series of simpler steps \citep{lehmann1998theory}. Moreover, hierarchical Bayesian modelling  may be seen as a way to account for uncertainty about the hyperparameters. 

A typical hierarchical model consists of a data level, a process level, and a prior level. The data layer relates observed measurements to latent variables by means of the likelihood. The process layer encodes the underlying scientific or generative mechanism, and the prior layer captures uncertainty about the model parameters and hyperparameters. This modular structure not only enhances interpretability but also facilitates the construction of complex models in a principled way. Hierarchical models are a central tool in modern Bayesian analysis, offering both conceptual clarity and computational flexibility for modeling structured, heterogeneous, or incomplete data.

Hierarchical models are particularly well suited for capturing complex data structures, such as those arising from nested populations, heterogeneous data sources with varying noise levels, or incomplete and biased measurements. Instead of compressing all sources of variation into a single layer, hierarchical modeling enables the decomposition of the problem into coherent, interpretable components. For instance, individual observations \( y_i \) may depend on latent parameters \( \theta_i \), which themselves are drawn from a common population distribution characterized by hyperparameters \( \mu \) and \( \tau^2 \). This framework supports partial pooling, balancing information from individual units with information shared across groups. As a result, hierarchical models provide a principled approach to borrowing strength across related units while preserving the ability to capture unit-specific effects.

\subsection{The impact of adding stages to the hierarchical model}

Let us represent a hierarchical Bayesian model specification as follows:

\[
 \boldsymbol{Y}\mid\boldsymbol{\theta}\sim p(\boldsymbol{y}\mid\boldsymbol{\theta}), \qquad
 \boldsymbol{\Theta}\mid\boldsymbol{\lambda}\sim p(\boldsymbol{\theta}\mid\boldsymbol{\lambda}), \qquad
 \boldsymbol{\Lambda}\sim \psi(\boldsymbol{\lambda}).
\]

Thus, when formulating this kind of model, questions regarding  the impact that the stages of the hierarchy have on each other emerge. In particular, it is of interest to assess how different the posterior and prior distributions of all model unknowns (parameters and hyperparameters) are. There is a generalized belief that parameters deeper in the hierarchy have a smaller impact on inference. Consequently, adding more stages or layers to the model has a small effect on the posterior and, therefore, there is less concern about choosing \(\psi(\boldsymbol{\lambda})\). A formal result addressing this phenomenon via the Kullback-Leibler divergence is found in \cite{lehmann1998theory} and is presented without proof.

For the hierarchical model presented above, it follows that:

\[
K\big(p(\boldsymbol{\lambda}\mid\boldsymbol{y}),\psi(\boldsymbol{\lambda})\big)<K\big(p(\boldsymbol{\theta}\mid\boldsymbol{y}),p(\boldsymbol{\theta)}\big),
\]

where \(K(\cdot,\cdot)\) is the Kullback-Leibler divergence, and

\[
p(\boldsymbol{\theta}\mid\boldsymbol{y})=\frac{p({\boldsymbol{y}}\mid\boldsymbol{\theta})\,p(\boldsymbol{\theta})}{p(\boldsymbol{y})},\qquad
p(\boldsymbol{\lambda}\mid\boldsymbol{y})=\frac{\int_\Theta p({\boldsymbol{y}}\mid\boldsymbol{\theta})\,p(\boldsymbol{\theta}\mid\boldsymbol{\lambda})\,\psi(\boldsymbol{\lambda})\,\textsf{d}\theta}{p(\boldsymbol{y})}.
 \]

Therefore, a common specification for \(\psi(\boldsymbol{\lambda})\) is a ``flat'' distribution (e.g., the Lebesgue measure of the support) which may be proper or improper. In addition to its simplicity, in certain problems. For example, inferring the multivariate normal mean, such a specification has another advantage, it leads to priors \(p(\boldsymbol{\theta})\) with heavier tails, leading to robust inference \citep{berger1990subjective}. 

\subsection{Example: Hierarchical Normal model}

To illustrate the structure of a hierarchical model, consider the following three-level normal formulation. At the first level, we model the observed data $y_1, \ldots, y_n$ as conditionally independent given latent parameters $\theta_1, \ldots, \theta_n$, with $y_i \mid \theta_i \overset{\text{ind.}}{\sim} \textsf{N}(\theta_i, \sigma^2)$ for $i = 1, \ldots, n$. This level captures measurement noise or individual-level variation, where $\sigma^2$ represents the observation-level variance.

At the second level, the latent parameters $\theta_i$ are assumed to arise from a common population distribution, $\theta_i \mid \mu, \tau^2 \overset{\text{i.i.d.}}{\sim} \textsf{N}(\mu, \tau^2)$, for $i = 1, \ldots, n$. This layer induces dependence among the $\theta_i$'s and enables partial pooling across units. The parameters $\mu$ and $\tau^2$ represent the overall mean and variance of the latent effects, respectively, and encode the population-level structure.

At the third level, we complete the hierarchy by treating $\mu$, $\tau^2$, and $\sigma^2$ as unknown and assigning them prior distributions, $\sigma^2, \mu, \tau^2 \sim p(\sigma^2, \mu, \tau^2)$. This prior layer expresses uncertainty about the group-level parameters and enables full Bayesian inference across all levels of the model. Such a formulation is flexible and widely applicable, particularly in settings involving exchangeability, nested variation, or multi-source data.

This hierarchical structure naturally induces shrinkage of the individual parameters \( \theta_i \) toward the group-level mean \( \mu \). Such shrinkage arises from the pooling of information across units and reflects the trade-off between individual-level fidelity and population-level regularization. This phenomenon is closely related to the James--Stein paradox, which shows that for \( n \geq 3 \), pooled estimators can dominate the maximum likelihood estimators in terms of mean squared error, even when the observations are independent.

Under the model given above,  the full joint posterior distribution is given by
\[
p(\theta_1, \ldots, \theta_n, \sigma^2, \mu, \tau^2 \mid \boldsymbol{y}) \propto 
\prod_{i=1}^n p(y_i \mid \theta_i, \sigma^2) \cdot \prod_{i=1}^n p(\theta_i \mid \mu, \tau^2) \cdot p(\sigma^2,\mu, \tau^2).
\]
This distribution reflects uncertainty across all levels of the model, including individual effects \( \theta_i \), group-level hyperparameters \( \mu \) and \( \tau^2 \), and observation-level variance \( \sigma^2 \).

However, inference in this model is analytically intractable, as it involves integrating over a high-dimensional, nested posterior distribution with no closed-form expression. The presence of multiple layers of dependency complicates marginalization and conditioning, particularly for posterior summaries or predictions. Consequently, Bayesian inference in hierarchical models typically relies on computational approaches such as MCMC, which enable sampling from the joint posterior and facilitate estimation of posterior means, variances, and credible intervals for all parameters of interest. These methods make hierarchical modeling not only conceptually appealing but also practically feasible.

It is worth mentioning that an alternative approach to account for hyperparameters uncertainty is the so-called empirical Bayes, which is based on estimating them using the observed data. It is achieved by obtaining the marginal likelihood (integrating out the model parameters), which yields the distribution of data given the hyperparameters \citep{lehmann1998theory}. Because of this procedure, the empirical Bayes approach falls outside the formal Bayesian paradigm, which has brought several critics, for example, the words of Dennis Victor Lindley ``no one is less Bayesian than en empirical Bayesian''. However, in certain problems, it can be used as an approach to hierarchical Bayes. Moreover, it yields estimators with Bayesian and frequentist desirable properties, one of the reasons for this phenomena is that these estimators tend to be robust against prior misspecification, besides, there are models for which point estimators under empirical are hierarchical Bayes are the same (\citealt{lehmann1998theory} and \citealt{ghosh1992hierarchical}). 

\section{Why adopting a Bayesian perspective?}

There are several compelling reasons to adopt a Bayesian perspective. Bayesian methods ensure admissibility under common loss functions, naturally accommodate exchangeability, and strictly adhere to the likelihood principle. They provide a coherent framework for hierarchical modeling and borrowing strength across related units. With advances in MCMC and simulation-based inference, Bayesian computation has become efficient and often simpler than frequentist alternatives for complex models. Moreover, treating both data and parameters as random variables simplifies the handling of missing or censored data via techniques like data augmentation.

\subsection{Reconciling Bayesian and frequentist perspectives}

In both, Bayesian and frequentist frameworks, statistical procedures are evaluated in terms of their performance under a specified loss function. Given a parameter \( \theta \) and an estimator \( h(\boldsymbol{y}) \), the loss \( L(\theta, h(\boldsymbol{y})) \) quantifies the cost of using \( h(\boldsymbol{y}) \) to estimate \( \theta \) based on the observed data. In the frequentist paradigm, this performance is typically assessed through the \emph{frequentist risk}, defined as
\[
R(\theta, h) = \textsf{E}_{\boldsymbol{y} \mid \theta} \left[ L(\theta, h(\boldsymbol{y})) \right] 
= \int_{\mathcal{Y}} L(\theta, h(\boldsymbol{y})) \, p(\boldsymbol{y} \mid \theta) \, \textsf{d}\boldsymbol{y},
\]
which represents the expected loss over repeated samples drawn from the data-generating process, conditional on a fixed value of \( \theta \).

A limitation of the frequentist perspective is that it treats the parameter \( \theta \) as fixed but unknown, making it unclear how to evaluate an estimator's overall performance when \( \theta \) is not observed. A natural Bayesian response is to average the risk over the parameter space using the prior distribution \( p(\theta) \), leading to the \emph{Bayes risk}:
\[
r(h, \pi) = \textsf{E}_\theta \left[ R(\theta, h) \right] = \int_\Theta R(\theta, h) \, p(\theta) \, \textsf{d}\theta 
= \int_{\Theta} \int_{\mathcal{Y}} L(\theta, h(\boldsymbol{y})) \, p(\boldsymbol{y} \mid \theta) \, p(\theta) \, \textsf{d}\boldsymbol{y} \, \textsf{d}\theta.
\]
This quantity reflects the expected loss of an estimator, weighted by the prior plausibility of each parameter value. Rather than emphasizing worst-case scenarios, Bayes risk provides a principled average-case criterion for comparing estimators under prior uncertainty.

From a Bayesian perspective, however, the primary focus is not on repeated sampling, but on the data actually observed. This leads naturally to the concept of \emph{posterior expected loss}, defined as
\[
\rho_{h, \pi}(\boldsymbol{y}) = \textsf{E}_{\theta \mid \boldsymbol{y}} \left[ L(\theta, h(\boldsymbol{y})) \right]
= \int_\Theta L(\theta, h(\boldsymbol{y})) \, p(\theta \mid \boldsymbol{y}) \, \textsf{d}\theta,
\]
which averages the loss with respect to the posterior distribution of \( \theta \) given the observed data \( \boldsymbol{y} \). The estimator that minimizes this posterior loss is the Bayes rule for the specific sample and represents the optimal decision under the Bayesian framework.

While the posterior expected loss focuses on performance conditional on the observed data, one may still ask how an estimator performs on average across repeated datasets. Within the Bayesian paradigm, this average is taken over the marginal distribution \( p(\boldsymbol{y}) \), which integrates over parameter uncertainty, rather than over the sampling distribution \( p(\boldsymbol{y} \mid \theta) \). This leads to the \emph{expected posterior loss}:
\[
\textsf{E}_{\boldsymbol{y}} \left[ \rho_{h, \pi}(\boldsymbol{y}) \right] 
= \int_{\mathcal{Y}} \rho_{h, \pi}(\boldsymbol{y}) \, p(\boldsymbol{y}) \, \textsf{d}\boldsymbol{y}
= \int_{\mathcal{Y}} \int_{\Theta} L(\theta, h(\boldsymbol{y})) \, p(\theta \mid \boldsymbol{y}) \, p(\boldsymbol{y}) \, \textsf{d}\theta \, \textsf{d}\boldsymbol{y}.
\]
This formulation evaluates the average posterior loss across all possible datasets, while preserving the Bayesian principle of conditioning on the data actually observed when making decisions.

Together, these formulations demonstrate that although frequentist and Bayesian approaches differ in interpretation, they are grounded in a common formal structure based on loss functions, expectations, and probability models. The divergence lies not in mathematical inconsistency but in philosophical orientation: frequentists average over data given a fixed parameter, while Bayesians average over parameters given observed data. Far from being incompatible, the two perspectives offer complementary insights into the evaluation and construction of statistical procedures.

\subsection{Admissibility}

A central question in decision theory is how to compare competing estimators in a principled and meaningful way. One of the most fundamental criteria is \emph{admissibility}. An estimator \( h_2 \) is said to be \emph{inadmissible} if there exists another estimator \( h_1 \) such that the risk function satisfies \( R(\theta, h_1) \leq R(\theta, h_2) \) for all \( \theta \in \Theta \), with strict inequality for at least one value of \( \theta \in \Theta \). In such a case, \( h_1 \) is said to \emph{dominate} \( h_2 \), and the inadmissible estimator \( h_2 \) should be discarded in favor of the better-performing alternative. Not surprisingly, many commonly used procedures turn out to be inadmissible, even under standard loss functions.

A well-known example of inadmissibility is provided by the \emph{James-Stein estimator}, which illustrates a fundamental distinction between Bayesian and frequentist approaches to estimation. Consider the problem of estimating a multivariate mean based on a single observation \( \boldsymbol{y} \sim \textsf{N}_d(\boldsymbol{\theta}, \mathbf{I}) \). Under squared error loss, the maximum likelihood estimator (MLE) is \( \hat{\boldsymbol{\theta}} = \boldsymbol{y} \). However, in dimensions \( d > 2 \), this estimator is inadmissible: there exists an alternative estimator with strictly smaller risk. One such dominating estimator is the James--Stein estimator,
\[
\tilde{\boldsymbol{\theta}}= \left( 1 - \frac{d - 2}{\boldsymbol{y}^\top \boldsymbol{y}} \right) \boldsymbol{y},
\]
which applies a data-dependent shrinkage factor that pulls the estimate toward the origin. This result is surprising because it implies that shrinking toward a common target, even in the absence of specific prior information, leads to uniformly better performance in terms of expected loss.

The inadmissibility of the MLE can be demonstrated by comparing its risk to that of the James--Stein estimator. Under squared error loss, the risk of the MLE is constant and equal to \( d \), since
\[
\textsf{E}_{\boldsymbol{y} \mid \boldsymbol{\theta}} \big[ \lVert \boldsymbol{\theta} - \boldsymbol{y} \rVert^2 \big] = \textsf{E}\big[ \, \chi^2_d \, \big] = d,
\]
because \( \boldsymbol{y} - \boldsymbol{\theta} \sim \textsf{N}_d(\boldsymbol{0}, \mathbf{I}) \), and its squared norm follows a chi-squared distribution with \( d \) degrees of freedom. In contrast, the James--Stein estimator achieves strictly smaller risk:
\[
\textsf{E}_{\boldsymbol{y} \mid \boldsymbol{\theta}} \big[ \lVert \boldsymbol{\theta} - \tilde{\boldsymbol{\theta}} \rVert^2 \big] 
= d - \textsf{E}_{\boldsymbol{y} \mid \boldsymbol{\theta}} \left[ \frac{(d-2)^2}{\boldsymbol{y}^\top \boldsymbol{y}} \right] < d.
\]
The second term is strictly positive for \( d > 2 \), implying that the James--Stein estimator uniformly dominates the MLE in terms of expected loss.

The James–Stein result is fully consistent with decision-theoretic reasoning and admits a natural Bayesian interpretation. In particular, the James–Stein estimator arises from the hierarchical model $y_i \mid \theta_i \sim \textsf{N}(\theta_i, 1)$, with $\theta_i \overset{\text{i.i.d.}}{\sim} \textsf{N}(0, \tau^2)$. When the prior variance $\tau^2$ is estimated from the data, the resulting empirical Bayes estimator closely resembles the James–Stein shrinkage rule. This formulation captures the Bayesian principle of \emph{borrowing strength} across components: information is shared across the $\theta_i$'s through their common prior, allowing individual estimates to benefit from the full dataset. The result is partial pooling, which balances individual variability with population-level regularization, thereby reducing total risk.

More broadly, admissibility results offer strong theoretical support for the Bayesian framework. Under standard regularity conditions (such as a well-specified parameter space, continuity of the loss function, and integrability of the risk) every Bayes estimator with finite Bayes risk is admissible. Conversely, every admissible estimator corresponds to a Bayes estimator under some (possibly improper) prior. These dual results reveal a deep connection between Bayesian optimality and admissibility, providing a compelling justification for Bayesian procedures even from a frequentist standpoint. A rigorous exposition of these results is given in \cite{robert2007bayesian}. This example underscores a key insight of Bayesian modeling: hierarchical priors can yield estimators that dominate standard methods, reinforcing the role of admissibility as a natural bridge between Bayesian and frequentist paradigms.

\subsection{Exchangeability}

A compelling rationale for Bayesian inference lies in the concept of \emph{exchangeability}. While admissibility links Bayesian procedures to decision-theoretic optimality, exchangeability offers a probabilistic foundation for modeling uncertainty in sequences of observations, especially when independence cannot be assumed. A sequence \( y_1, y_2, \ldots \) is said to be \emph{exchangeable} if its joint distribution is invariant under permutations. Formally, for any \( n \in \mathbb{N} \) and any permutation \( \sigma \) of \( \{1, \ldots, n\} \), exchangeability holds if
\[
p(y_1, \ldots, y_n) = p(y_{\sigma(1)}, \ldots, y_{\sigma(n)}).
\]
That is, the order of observations conveys no information. Although every i.i.d.\ sequence is exchangeable, the converse is not true. Exchangeability is strictly more general. For example, a jointly Gaussian sequence with zero mean, unit variance, and constant correlation \( \rho \in (0,1) \) is exchangeable but not i.i.d.

The foundational role of exchangeability in Bayesian modeling is captured by \emph{de Finetti's theorem}. Suppose that for each $n$, the joint distribution $p(y_1, \ldots, y_n)$ is exchangeable. Then there exists a latent parameter $\theta$, a sampling distribution $p(y \mid \theta)$, and a prior distribution $p(\theta)$ such that the joint distribution admits the representation
$$
p(y_1, \ldots, y_n) = \int \prod_{i=1}^n p(y_i \mid \theta) \, p(\theta) \, \textsf{d}\theta.
$$
The specific forms of $p(\theta)$ and $p(y \mid \theta)$ depend on the structure of the exchangeable model. In the binary case, the prior $p(\theta)$ encodes beliefs about the limiting frequency $\lim_{n \to \infty} \frac{1}{n} \sum_{i=1}^n y_i$. More generally, for real-valued observations, it represents beliefs about functionals such as $\lim_{n \to \infty} \frac{1}{n} \sum_{i=1}^n \mathbb{I}(y_i \leq c)$, for all $c \in \mathcal{Y}$. This representation justifies the use of hierarchical Bayesian models: any exchangeable sequence can be interpreted as a mixture of conditionally i.i.d.\ observations given an underlying latent structure.

Together with admissibility, exchangeability reinforces the Bayesian framework not only as a coherent decision-theoretic system but also as a natural approach to modeling data-generating processes. It provides justification for hierarchical modeling, and unifies statistical reasoning under uncertainty. Thus, the Bayesian perspective emerges as a principled framework rooted in both optimality and foundational representation theorems.

\subsection{Some drawbacks of frequentist inference}

Under the frequentist approach to statistical inference, the unknown parameter is treated as a fixed (i.e., nonrandom) quantity. Inference is therefore framed in terms of procedures whose long run behavior is evaluated under repeated sampling, rather than in terms of probability statements about the parameter for the realized data set \cite{lecoutre2022significance}. In particular, frequentist confidence intervals are typically justified by their coverage properties under hypothetical replications of the experiment, which may be hard to interpret or defend when attention is restricted to the actual data at hand.

This concern becomes more acute in designs where additional sources of randomness enter through the sampling rule or through post data randomization, because the nominal frequentist guarantees are then averages over events that did not occur. Put differently, the unconditional frequentist probability statement is taken with respect to an experiment that includes random elements that may be irrelevant once the realized outcome is observed. Conditional frequentist inference, which conditions on ancillary statistics or on aspects of the sampling mechanism, is one response to this difficulty, although in general there is no universally accepted conditioning principle.

Moreover, evaluating the performance of a frequentist inference procedure often implies the computation of expected values (e.g., bias or standard errors), which are obtained by integrating over the sample space (of possible data), this leads to a non-pleasant feature, namely that these quantities do not inform about the performance of the inferential method for the particular data set being used. The following stylized examples are useful to further illustrate these shortcomings. 

\subsubsection{Example 1: Cox}
Assume that
$X_1,\ldots,X_n \mid \mu \overset{\text{ind}}{\sim} \textsf{N}(\mu,\sigma^2)$,
with $\sigma^2$ known. To estimate $\mu$, toss a fair coin and choose the sample size according to the outcome. If the result is head, take $n=2$, and if it is tail, take $n=1000$. Consider the estimator $\bar{X}$. It is unbiased since
\begin{align*}
\textsf{E}(\bar{X})
&=
\frac{1}{2}\,\textsf{E}\!\left(\bar{X} \mid n=2\right)
+
\frac{1}{2}\,\textsf{E}\!\left(\bar{X} \mid n=1000\right)
=
\mu .
\end{align*}
Unconditionally, its variance is
\begin{align*}
\textsf{Var}(\bar{X})
&=
\frac{1}{2}\,\textsf{Var}\!\left(\bar{X} \mid n=2\right)
+
\frac{1}{2}\,\textsf{Var}\!\left(\bar{X} \mid n=1000\right) \\
&=
\frac{1}{2}\,\frac{\sigma^2}{2}
+
\frac{1}{2}\,\frac{\sigma^2}{1000}
=
\frac{\sigma^2}{4}+\frac{\sigma^2}{2000}
\approx
\frac{\sigma^2}{4},
\end{align*}
recall that $\sigma^2$ is known. If the observed outcome of the coin toss is tail, hence $n=1000$, the relevant uncertainty for the realized estimator is instead $\textsf{Var}(\bar{X} \mid n=1000)=\sigma^2/1000$, so using $\sigma^2\approx
\frac{\sigma^2}{4}$ is not useful. This illustrates how the unconditional variance, which averages over the sampling rule, can be misleading for the actual data set.

It is worth mentioning that frequentist statisticians like Ronald Fisher were aware of this phenomenon, he suggested conditional inference, using an ancillary statistic for such purpose, in this case it would be sample size. 

A related issue appears when randomization is introduced after data collection. Suppose $\sigma^2$ is known and we wish to estimate $\mu$. Toss a fair coin, if it is head take $\hat{\mu}=X_1$, and if it is tail take $\hat{\mu}=(X_1+X_2)/2$. Then
\begin{align*}
\textsf{E}(\hat{\mu})
&=
\frac{1}{2}\,\textsf{E}\!\left(X_1 \mid \text{head}\right)
+
\frac{1}{2}\,\textsf{E}\!\left(\frac{X_1+X_2}{2} \,\middle|\, \text{tail}\right)
=
\frac{1}{2}\mu+\frac{1}{2}\mu
=
\mu ,
\end{align*}
and
\begin{align*}
\textsf{Var}(\hat{\mu})
&=
\frac{1}{2}\sigma^2+\frac{1}{2}\frac{\sigma^2}{2}
=
\frac{3}{4}\sigma^2.
\end{align*}
If the realized outcome is tail, the estimator becomes $(X_1+X_2)/2$ and the relevant variance is $\sigma^2/2$, making the unconditional statement $\textsf{Var}(\hat{\mu})=3\sigma^2/4$ difficult to interpret for the realized experiment. \hfill$\square$

\subsubsection{Example 2: Welch}
Assume that
$X_1,X_2 \mid \theta \overset{\text{ind}}{\sim} \textsf{U}(\theta-\tfrac{1}{2},\theta+\tfrac{1}{2})$.
A $95\%$ confidence interval for $\theta$ can be defined through an interval of the form
\[
\theta \in \frac{X_1+X_2}{2} \pm d(X_1,X_2),
\]
with $d>0$ suitably chosen so that
\[
\textsf{P}\!\left(
\frac{X_1+X_2}{2}-d(X_1,X_2)
<
\theta
<
\frac{X_1+X_2}{2}+d(X_1,X_2)
\right)
=
0.95.
\]
Now suppose that we observe $X_1=2$ and $X_2=1$. In this case we know that
$\theta=(X_1+X_2)/2=1.5$,
so it becomes a certain event that $\theta \in (\bar{X}-d,\bar{X}+d)$ for any $d>0$. From the perspective of conditional inference, one can condition on the ancillary statistic $X_1-X_2=1$, whose distribution does not depend on $\theta$ in this location model. This example shows that the nominal unconditional coverage can differ substantially from the conditional behavior relevant for the observed ancillary information.

For these two problems, it turns out that the conditional frequentist estimators recommended by Fisher, agree with those obtained under an objective Bayesian formulation (based on Jeffreys prior), which is the typical case for location and scale parameters \citep{GhoshEtAl2007}. 

These examples highlight a general drawback of unconditional frequentist reasoning in the presence of ancillary information or additional randomization, namely that guarantees based on averaging over unobserved outcomes may not reflect the inferential uncertainty for the realized data set. See, for instance, \cite{lecoutre2022significance}, \cite{GhoshEtAl2007}, and \cite{ghosh2011objective} for discussions of these issues and of conditional frequentist alternatives. \hfill$\square$

\subsection{The likelihood principle}

Another foundational argument in favor of the Bayesian framework is the \emph{likelihood principle}. This principle asserts that all relevant information for making inferences about the parameter $\theta$, once the data $\boldsymbol{y}$ have been observed, is fully contained in the likelihood function $p(\boldsymbol{y} \mid \theta)$. Stated anther way, inference shall be entirely based on the likelihood, ignoring the sample space. In particular, if two datasets $\boldsymbol{y}$ and $\boldsymbol{y}'$ yield proportional likelihood functions, i.e., $p(\boldsymbol{y} \mid \theta) \propto p(\boldsymbol{y}' \mid \theta)$, then any inference about $\theta$ should be identical under both datasets. 

While not a theorem, the likelihood principle follows naturally from the Bayesian perspective, where inference is conditioned solely on the observed data and prior beliefs. Moreover, the posterior depends on the data only through the likelihood. Unlike frequentist methods, which may incorporate the sampling distribution beyond the realized data, Bayesian inference adheres to this principle by design. In this way, the likelihood principle complements the ideas of admissibility and exchangeability: it emphasizes the coherence of Bayesian updating and reinforces the notion that inference should depend only on the information actually observed, rather than on hypothetical repetitions or alternative experimental designs.

A classic illustration of the likelihood principle involves comparing binomial and negative binomial sampling schemes. Consider two independent experiments aimed at estimating a Bernoulli success probability $\theta$. In the first, the researchers fix the number of trials at $n = 12$ and observe $y_1 = 3$ successes. The likelihood for this binomial experiment is proportional to $\binom{12}{3} \theta^3 (1 - \theta)^9$. In the second experiment, the researchers fix the number of successes at 3 and continue sampling until this condition is met, requiring $y_2 = 12$ total trials. This leads to a negative binomial likelihood, proportional to $\binom{11}{2} \theta^3 (1 - \theta)^9$. Since these likelihoods differ only by a multiplicative constant, they are proportional, and the likelihood principle implies that any inference about $\theta$ should be identical in both cases.

Inference procedures that respect the likelihood principle, such as maximum likelihood estimation and Bayesian inference under proper priors, produce identical conclusions when the likelihood functions are proportional. In the example at hand, both the binomial and negative binomial experiments yield the same likelihood for $\theta$, and hence the maximum likelihood estimate is $\hat{\theta} = 1/4$ in both cases. However, frequentist hypothesis testing does not conform to this principle. Consider testing $H_0: \theta \geq 1/2$ versus $H_1: \theta < 1/2$. Under the binomial sampling scheme, the $p$-value is $\Pr(y_1 \leq 3 \mid \theta = 1/2) \approx 0.07300$, while under the negative binomial model, it is $\Pr(y_2 \geq 12 \mid \theta = 1/2) \approx 0.03271$. Despite both experiments providing the same inferential content about $\theta$, the conclusions differ because the $p$-value depends not only on the observed outcome but also on the sampling plan. This dependence on unobserved data outcomes violates the likelihood principle and highlights a key philosophical divergence between Bayesian and frequentist paradigms.

As this example illustrates, a key implication of the likelihood principle is that, provided the stopping rule does not depend on the unknown parameters of the model, statistical inference should not be influenced by how or when the data collection process was terminated. In other words, inference should depend solely on the observed data through the likelihood function, not on the experimental design that generated it. 

Furthermore, revisiting Welch's and Box's examples, it has been recommended that a sound inferential procedure should not only be based on conditioning (as motivated in these two examples), but also on minimal sufficient statistics \citep{GhoshEtAl2007}. There is a theorem due to Birnbaum, that connects these two properties, namely conditional inference (CP) and the use of minimal sufficient statistics (MP) with the likelihood principle because it shows that to achieve CP and MP, the likelihood principle must be satisfied, so most frequentist procedures do not meet these properties.  

\subsection{Practical considerations}

A further advantage of the Bayesian framework is its natural accommodation of hierarchical structures and the principle of borrowing strength across related units. While hierarchical models can also be formulated in frequentist settings, the Bayesian approach offers a more coherent and interpretable foundation by treating all unknown quantities as random variables. Moreover, the development of modern computational techniques such as MCMC has made fully Bayesian inference computationally tractable even in high-dimensional or otherwise complex models. In many cases, Bayesian computation is now more flexible than classical alternatives, particularly when standard asymptotic approximations fail. Furthermore, in section xxxx we discussed how a set of axioms led to the Bayesian approach, which provides a strong theoretical underpinning. Also, Bayesian inference is well-suited to high-dimensional problems, a topic of contemporary interest, and hierarchical Bayesian modeling permits to accommodate complex data structures in a natural way. Besides, in many applied problems there is enough prior information and knowledge that should be incorporated into the inference process, the Bayesian approach can handle this situation via subjective priors. Bayesian ideas and measures are easy to interpret, and thus, they are easy to communicate. Moreover, accounting for uncertainty can be performed in a conceptually simple and principled way, hence the saying “uncertainty is for free”. Examples of this particular advantage in real-life problems may be found in \citet{martinez2017joint} and \citet{martinez2018bibi}. Finally, the Bayesian treatment of parameters as stochastic facilitates principled handling of missing or censored data through data augmentation schemes, further highlighting the adaptability of Bayesian approaches to real-world modeling challenges.

\section{Challenges in Bayesian inference}

While the Bayesian framework offers conceptual coherence and practical flexibility, it is not without challenges. The most prominent difficulties involve the elicitation of prior distributions, especially in complex or high-dimensional models, and the computational demands of deriving posterior distributions and associated summaries when conjugacy is unavailable. These difficulties are further compounded when decision-making involves non-standard utility functions. Addressing these challenges is essential for the effective and responsible application of Bayesian methods in real-world settings.

\subsection{Prior elicitation}

In the Bayesian framework, a statistical model consists of both a likelihood function and a prior distribution. The prior is not a secondary element but a fundamental component that warrants the same level of scrutiny as the likelihood. This dual structure is sometimes viewed as a limitation, since different researchers may adopt different priors and thus reach differing conclusions. While such subjectivity can be concerning, its impact is typically attenuated in large samples, where the data tend to dominate the posterior. Nevertheless, in small-sample settings, poorly specified priors can substantially distort inference and lead to misleading conclusions.

Eliciting meaningful prior information is undoubtedly challenging. Yet, when carefully specified, informative priors can significantly enhance the accuracy and stability of inference, particularly in data-scarce settings. Moreover, the idea that only Bayesian methods rely on prior assumptions is misleading. All statistical models implicitly encode prior beliefs through structural and parametric choices. This is especially evident in the adoption of penalized likelihood approaches, where regularization terms act as implicit priors. In this light, the distinction between “modeling” through the likelihood and “prior elicitation” is artificial, since both represent assumptions about the data-generating process and deserve equal methodological attention.

A common distinction in Bayesian analysis lies between \textit{subjective} and \textit{objective} approaches. In subjective Bayes, the prior is explicitly derived from expert knowledge (often external to the analyst) and specified as a proper probability distribution. In contrast, objective Bayes treats the prior as a formal tool to enable the use of Bayes’ theorem. It may be improper (i.e., not integrable to one), so long as the resulting posterior is proper. Objective priors are typically chosen to satisfy criteria such as invariance, minimal informativeness, or desirable frequentist properties like correct coverage. In practice, a balanced strategy is often preferred: incorporate prior information when it is available, but rely on well-motivated default priors in high-dimensional or data-limited settings where elicitation is impractical.

\subsection{Subjective Bayes}

In practice, prior elicitation is a challenging task because it requires specifying an entire probability distribution. While domain experts may be able to articulate beliefs about specific features of a distribution (such as a few quantiles or moments) they are rarely equipped to provide the level of detail necessary to fully characterize a prior. As a result, analysts often restrict attention to parametric families with a small number of interpretable parameters.

Even with modern simulation-based computational tools, the choice of prior distributions is frequently guided by computational convenience. In particular, families that yield conditionally conjugate priors are preferred, as they facilitate efficient posterior computation. Once a parametric family is selected, the elicitation process reduces to identifying enough summary measures (e.g., quantiles or moments) to solve for the hyperparameters.

As previously mentioned, an alternative strategy for prior specification is the \emph{empirical Bayes} approach. Rather than specifying a prior distribution entirely \emph{a priori}, empirical Bayes methods estimate the hyperparameters directly from the observed data, thereby allowing the data to partially inform the prior. Recall that this requires integrating out the model parameters to obtain the marginal likelihood and then maximizing it with respect to the hyperparameters, a procedure also known as \emph{Type II maximum likelihood}. In practice, however, this full optimization is often computationally intensive and rarely implemented in full generality. Instead, empirical Bayes procedures typically rely on simpler, data-driven estimates (such as matching sample moments or using quantiles) to determine the location and scale of the prior distribution.

As with direct elicitation, empirical Bayes methods often rely on information about the observables to guide the choice of hyperparameters, inferring what such information implies for the prior distribution. A key limitation of this approach is that it uses the same data both to estimate the prior and to conduct inference, thereby blurring the conceptual separation between prior information and observed evidence. This double use of data can undermine some of the formal optimality guarantees of fully Bayesian procedures. Nevertheless, empirical Bayes remains a practical and widely adopted strategy, particularly in settings where full prior elicitation is difficult or impractical.

\subsection{Objective Bayes: Laplace’s indifference principle}

A classical justification for the use of \emph{default priors} (priors chosen not to reflect subjective beliefs but to enable general-purpose inference) is given by \emph{Laplace’s principle of indifference}. This principle states that when \( K \) mutually exclusive outcomes are indistinguishable except for their labels, each should be assigned equal probability \( 1/K \). While the principle is well-defined in finite parameter spaces, its extension to countably infinite or continuous spaces can lead to \emph{improper priors}, i.e., functions that do not integrate to one. Such priors may still be useful, but they require caution: the resulting posterior distribution is proper if and only if
\[
\int_\Theta p(\boldsymbol{y} \mid \theta) \, p(\theta) \, \textsf{d}\theta < \infty,
\]
a condition that always holds when \( p(\theta) \) is proper. For improper priors, however, this must be checked explicitly to ensure that the posterior is well-defined and supports valid inference.

\subsubsection{Example: Normal likelihood}

Assume that $y_i \overset{\text{i.i.d.}}{\sim} \textsf{N}(\theta, 1)$, for $i = 1, \ldots, n$. Applying Laplace's principle of indifference to $\theta$ leads to an improper prior \( p(\theta) \propto 1 \). Despite the prior being improper, the posterior is proper. Indeed, the likelihood is
$$
p(\boldsymbol{y} \mid \theta) \propto \exp\left\{ -\frac{1}{2} \sum_{i=1}^n (y_i - \theta)^2 \right\} =\exp\left\{ -\frac{n}{2}(\theta - \bar{y})^2 + \text{c} \right\},
$$
where $\bar{y} = \frac{1}{n} \sum_{i=1}^n y_i$ and c is a constant. The posterior is thus proportional to the kernel of a normal distribution, and we conclude that $\theta \mid \boldsymbol{y} \sim \textsf{N}\left( \bar{y}, \frac{1}{n} \right)$. This example illustrates that even when using an improper prior, the posterior can remain proper, provided the data supply enough information. \hfill$\square$

\subsubsection{Example: Bernoulli likelihood}

Consider a single Bernoulli observation $y \in \{0,1\}$ with success probability parameterized as $\theta = \frac{e^\phi}{1 + e^\phi}$, where $\phi \in \mathbb{R}$ is the log-odds. Applying Laplace's principle of indifference to $\phi$ suggests using the uniform prior $p(\phi) \propto 1$. This prior is sometimes adopted as a non-informative prior on the logit scale. The likelihood under this parameterization is
$$
p(y \mid \phi) = \left( \frac{\exp(\phi)}{1 + \exp(\phi)} \right)^y \left( 1 - \frac{1}{1 + \exp\{\phi\}} \right)^{1 - y} = \frac{\exp(y\phi)}{1 + \exp(\phi)}.
$$
We examine the normalization constant of the posterior:
$$
\int_{-\infty}^{\infty} \frac{\exp(y\phi)}{1 + \exp(\phi)} \, \textsf{d}\phi.
$$
If $y = 0$, the integrand reduces to $(1 + e^\phi)^{-1}$, which behaves like 1 as $\phi \to -\infty$, implying the integral diverges. If $y = 1$, the integrand becomes $e^\phi / (1 + e^\phi)$, which tends to 1 as $\phi \to \infty$, again leading to divergence. Thus, the posterior is improper for both $y = 0$ and $y = 1$. In contrast, if we place a uniform prior on the probability scale, i.e., $\theta \sim \textsf{Uniform}(0,1)$, then the posterior $p(\theta \mid y) \propto \theta^y (1 - \theta)^{1 - y}$ is properly normalized on $[0,1]$, since $\theta \mid y \sim \textsf{Beta}(1 + y, 2 - y)$.

This example underscores a fundamental limitation of Laplace’s principle of indifference in continuous parameter spaces: it offers no guidance on which parameterization should receive a uniform prior. Since a prior that is flat in one parameterization can be highly informative in another, the choice of scale or transformation becomes a critical and nontrivial modeling decision. This ambiguity challenges the very notion of a universally non-informative or default prior. \hfill$\square$

\subsection{Objective Bayes: Jeﬀreys prior}

The \textit{Jeffreys prior} is motivated by the goal of achieving invariance under reparameterization, a property that Laplace’s principle of indifference lacks in continuous parameter spaces. Jeffreys proposed defining the prior through the Fisher information, thereby capturing the intrinsic curvature of the parameter space induced by the likelihood. Formally, the Jeffreys prior is given by
$$
p_{\text{J}}(\boldsymbol{\theta}) \propto \left| \, \mathbf{I}(\boldsymbol{\theta}) \, \right|^{1/2},
$$
where $\mathbf{I}(\boldsymbol{\theta})$ is the Fisher information matrix with entries
$$
\mathbf{I}(\boldsymbol{\theta})_{i,j} = -\textsf{E}_{\boldsymbol{y} \mid \boldsymbol{\theta}} \left[ \frac{\partial^2}{\partial \theta_i \, \partial \theta_j} \log p(\boldsymbol{y} \mid \boldsymbol{\theta}) \right].
$$
This construction guarantees that the Jeffreys prior transforms appropriately under smooth reparameterizations, as the Fisher information naturally accounts for the Jacobian of the transformation. It thus offers a principled, parameterization-invariant default that aligns with the model’s likelihood structure.

In the case of standard parametric families, Jeffreys priors often admit simple closed-form expressions. For location families of the form $p(y \mid \theta) = p(y - \theta)$, where the data-generating process is invariant to shifts in the parameter, the Jeffreys prior is constant:
$$
p_{\text{J}}(\theta) \propto 1.
$$
This reflects uniform ignorance over the location parameter and corresponds to Laplace’s principle of indifference. For scale families of the form $p(y \mid \theta) = \frac{1}{\theta} f\left( \frac{y}{\theta} \right)$, the Jeffreys prior is inversely proportional to the scale:
$$
p_{\text{J}}(\theta) \propto \frac{1}{\theta},
$$
which is equivalent to assigning a uniform prior on the logarithmic scale of $\theta$. In both cases, the Jeffreys prior is improper, and it is essential to verify that the resulting posterior distribution is proper for valid inference. 

\subsubsection{Example: Binomial distribution}

Let $y \sim \textsf{Bin}(n, \theta)$. The log-likelihood function is
$$
\log p(y \mid \theta) = \log \binom{n}{y} + y \log \theta + (n - y) \log(1 - \theta).
$$
Differentiating twice with respect to $\theta$, we obtain
$$
- \frac{\textsf{d}^2}{\textsf{d}\theta^2} \log p(y \mid \theta) = \frac{y}{\theta^2} + \frac{n - y}{(1 - \theta)^2}.
$$
Taking the expectation with respect to the binomial sampling distribution yields the Fisher information:
$$
I(\theta) = \textsf{E}_{y \mid \theta} \left[ -\frac{\textsf{d}^2}{\textsf{d}\theta^2} \log p(y \mid \theta) \right] = \frac{n}{\theta(1 - \theta)}.
$$
Thus, the Jeffreys prior is
$$
p_{\text{J}}(\theta) \propto \sqrt{I(\theta)} = \theta^{-1/2}(1 - \theta)^{-1/2},
$$
which is the kernel of a $\textsf{Beta}(1/2, 1/2)$ distribution. This prior is proper and leads to a valid posterior distribution.

Now consider reparameterizing the binomial model in terms of the log-odds $\phi = \log\left( \frac{\theta}{1 - \theta} \right)$, so that $\theta = \frac{e^\phi}{1 + e^\phi}$. Applying the change-of-variable formula to the Jeffreys prior previously derived for $\theta$, we obtain the induced prior on $\phi$:
$$
p_{\text{J}}(\phi) = p_{\text{J}}(\theta) \cdot \left| \frac{\textsf{d}\theta}{\textsf{d}\phi} \right| 
\propto \left( \frac{e^\phi}{1 + e^\phi} \right)^{-1/2} \left( \frac{1}{1 + e^\phi} \right)^{-1/2} \cdot \frac{e^\phi}{(1 + e^\phi)^2} 
= \frac{e^{\phi/2}}{1 + e^\phi}.
$$
This expression is the Jeffreys prior on the log-odds scale. Notably, it agrees with the prior that results from applying Jeffreys’ rule directly to the log-likelihood expressed in terms of $\phi$,
$$
\log p(y \mid \phi) = y \phi - \log(1 + e^\phi),
$$
showing that the Jeffreys prior is invariant under smooth reparameterization. This invariance property reinforces its appeal as a default prior that adapts naturally to the geometry of the likelihood. \hfill$\square$

\subsubsection{Example: Negative Binomial distribution}

Let \( y \mid \theta \sim \textsf{NB}(r, \theta) \), where \( y = r, r+1, \ldots \) denotes the total number of trials required to observe \( r \) successes, and \( \theta \in (0,1) \) is the success probability. The log-likelihood function is:
\[
\log p(y \mid \theta) = \log \binom{y - 1}{r - 1} + r \log \theta + (y - r) \log(1 - \theta),
\]
whose second derivative with respect to \( \theta \) is
\[
- \frac{\textsf{d}^2}{\textsf{d}\theta^2} \log p(y \mid \theta) = \frac{r}{\theta^2} + \frac{y - r}{(1 - \theta)^2}.
\]
Taking expectation with respect to \( y \mid \theta \), using \( \textsf{E}(y\mid\theta) = \frac{r}{\theta} \), we obtain the Fisher information:
\[
I(\theta) = \textsf{E} \left[ -\frac{\textsf{d}^2}{\textsf{d}\theta^2} \log p(y \mid \theta) \right] = \frac{r}{\theta^2(1 - \theta)}.
\]
Thus, the Jeffreys prior becomes:
$$
p_{\text{J}}(\theta) \propto \sqrt{I(\theta)} = \left( \frac{r}{\theta^2 (1 - \theta)} \right)^{1/2}
\propto \theta^{-1} (1 - \theta)^{-1/2}.
$$
In this case, the Jeffreys prior is improper, since $\int_0^1 \theta^{-1}(1 - \theta)^{-1/2} \, \textsf{d}\theta$ diverges. 

Moreover, this Jeffreys prior differs from the one obtained in the binomial setting, despite the likelihoods being proportional for fixed outcomes. This underscores a key limitation: Jeffreys priors depend not only on the likelihood function but also on the underlying sampling scheme. As a result, Bayesian analyses using Jeffreys priors may violate the likelihood principle, yielding different inferences from data that are equally informative about the parameter. \hfill$\square$

\subsubsection{Example: Normal distribution}

Let $y_i \mid \theta, \phi \overset{\text{i.i.d.}}{\sim} \textsf{N}(\theta, \phi)$, for $i = 1, \ldots, n$, where $\phi = \sigma^2$ denotes the variance. Let $\boldsymbol{\theta} = (\theta, \phi)$ be the parameter vector. The log-likelihood function is
$$
\log p(\boldsymbol{y} \mid \theta, \phi) = -\frac{n}{2} \log(2\pi \phi) - \frac{1}{2\phi} \sum_{i=1}^n (y_i - \theta)^2.
$$
Taking derivatives and computing expectations with respect to the data distribution, we obtain the Fisher information matrix:
$$
\mathbf{I}(\boldsymbol{\theta}) =
\begin{pmatrix}
\frac{n}{\phi} & 0 \\
0 & \frac{n}{2\phi^2}
\end{pmatrix},
$$
whose determinant is $\left| \mathbf{I}(\boldsymbol{\theta}) \right| = \frac{n^2}{2\phi^3}$. Consequently, the Jeffreys prior is
$$
p_{\text{J}}(\boldsymbol{\theta}) \propto \left| \mathbf{I}(\boldsymbol{\theta}) \right|^{1/2} \propto \phi^{-3/2}.
$$

This prior is improper and sometimes exhibits undesirable behavior in multivariate settings due to its strong influence in the tails. As an alternative, the \emph{independence Jeffreys prior} assumes prior independence between $\theta$ and $\phi$, applying the Jeffreys rule separately to each component. Since the Jeffreys prior for $\theta$ is uniform and for $\phi$ is $\phi^{-1}$, the resulting independence Jeffreys prior is
$$
p_{\text{IJ}}(\boldsymbol{\theta}) \propto \frac{1}{\phi},
$$
which is widely used in practice due to its simplicity and better marginal behavior. However, it is important to note that independence Jeffreys priors are not invariant under transformations involving both parameters, and must be applied with caution. \hfill$\square$

\subsubsection{Barlett’s paradox}

Improper priors are widely used in Bayesian analysis and often work well for point estimation and interval construction, as they typically yield proper posteriors in regular models. However, their use becomes problematic in settings involving model comparison, particularly when models differ in dimensionality. In such cases, improper priors can distort Bayes factors and lead to paradoxical conclusions. A well-known example is \textit{Bartlett’s paradox}, which shows that using an improper prior under the alternative hypothesis can unjustifiably favor the null, regardless of the data.

Let $y_i \mid \theta \overset{\text{i.i.d.}}{\sim} \textsf{N}(\theta, 1)$, and consider testing the point null hypothesis $H_0: \theta = 0$ against the alternative $H_1: \theta \neq 0$. Under $H_1$, we assign a normal prior $\theta \sim \textsf{N}(0, \tau^2)$, centered at zero to ensure symmetry with the null. The marginal likelihood under the null is
$$
p(\boldsymbol{y} \mid H_0) = (2\pi)^{-n/2} \exp\left\{ -\frac{1}{2} \sum_{i=1}^n y_i^2 \right\},
$$
while under the alternative it becomes
$$
p(\boldsymbol{y} \mid H_1) = (2\pi)^{-n/2} \exp\left\{ -\frac{1}{2} \sum_{i=1}^n y_i^2 \right\} \cdot (n\tau^2 + 1)^{-1/2} \exp\left\{ \frac{n^2 \bar{y}^2}{2(1/\tau^2 + n)} \right\},
$$
where $\bar{y}=\frac{1}{n}\sum_{i=1}^n y_i$ is the sample mean. The Bayes factor comparing $H_1$ to $H_0$ is then
$$
\text{BF}_{1,0}(\tau^2, \bar{y}) = (n\tau^2 + 1)^{-1/2} \exp\left\{ \frac{n^2 \bar{y}^2}{2(1/\tau^2 + n)} \right\}.
$$
As $\tau^2 \to \infty$, this expression converges to zero, regardless of the observed data. Consequently, the posterior probability of the null tends to one. 

This behavior may initially seem paradoxical: when the sample mean $\bar{y}$ is large, we would naturally expect the Bayes factor to favor the alternative hypothesis $H_1$. Indeed, for fixed $\tau^2$, the Bayes factor increases with $\bar{y}^2$, and $\lim_{\bar{y} \to \infty} \text{BF}_{1,0}(\tau^2, \bar{y}) = \infty$. However, as the prior variance $\tau^2 \to \infty$—corresponding to a diffuse or improper prior under $H_1$—the Bayes factor tends to zero, regardless of the data. This apparent contradiction, where strong evidence is nullified by a flat prior, encapsulates \textit{Bartlett’s paradox}.

The key insight is that the order of limits matters: taking $\tau^2 \to \infty$ before observing the data leads to a degenerate Bayes factor that always favors the null. This arises because improper priors lack a normalizing constant, and Bayes factors are sensitive to such constants when comparing models of different dimensionality. In contrast, these constants cancel in posterior estimation, avoiding the issue.

Bartlett’s paradox highlights a broader point: priors can heavily influence Bayesian model selection, even with large samples. Using default or improper priors in hypothesis testing requires caution. In particular, testing point nulls against composite alternatives demands carefully calibrated proper priors to ensure that the Bayes factor accurately reflects the evidence. \hfill$\square$
\section{Discussion}

This work provided a structured introduction to the Bayesian paradigm, combining historical foundations, core principles, and analytical examples. It showed how prior beliefs and data are coherently integrated through Bayes’ theorem to produce posterior distributions that support estimation, uncertainty quantification, hypothesis testing, and prediction. Classical models such as the Normal–Normal, Binomial–Beta, and Poisson–Gamma families illustrated conjugate updates in closed form, while more general cases emphasized the flexibility of Bayesian inference, particularly when guided by loss-based decision rules. Asymptotic approximations, including Laplace expansions and the Bayesian Central Limit Theorem, further highlighted the theoretical coherence of the approach and its convergence to frequentist procedures in large samples. Foundational ideas such as admissibility, exchangeability, and the likelihood principle reinforced the philosophical consistency and probabilistic logic of the Bayesian framework.

Several essential components of modern Bayesian practice were not addressed in detail here. Posterior computation in high-dimensional or non-conjugate models frequently relies on simulation-based algorithms such as Markov Chain Monte Carlo, Hamiltonian Monte Carlo, and Variational Inference. Tools for model evaluation and comparison—including posterior predictive checks, residual diagnostics, and criteria such as DIC, WAIC, and leave-one-out cross-validation—play a crucial role in assessing model adequacy and guiding model selection. The formulation of prior distributions remains an active area of research, with increasing reliance on hierarchical, shrinkage, mixture-based, and nonparametric priors such as the Dirichlet and Pitman–Yor processes, which allow for flexible modeling of sparsity, heterogeneity, and latent structure.

This exposition also did not examine the use of Bayesian methods in several key areas of statistical modeling. These include linear models, generalized linear models, mixture models, dynamic models, spatial analysis, time series, survival models, longitudinal data, and the modeling of network structure. In political science, Bayesian approaches have proven particularly useful for studying legislative behavior, for example through ideal point estimation based on roll-call data. In each of these domains, the Bayesian framework offers a coherent foundation for capturing uncertainty, modeling dependence, borrowing strength across units, and improving predictive performance.

In conclusion, although the discussion here focused on foundational and analytical aspects, the practical effectiveness of Bayesian inference relies heavily on modern computational methods, robust diagnostic tools, and the capacity to adapt to increasingly complex modeling challenges. These elements are essential for the successful application of Bayesian statistics in contemporary scientific and applied research.




\bibliographystyle{apalike}
\bibliography{references}

\end{document}